\documentclass[lettersize,journal]{IEEEtran}
\usepackage{amsmath,amsfonts}
\usepackage{algorithmic}
\usepackage{algorithm}
\usepackage{array}
\usepackage[caption=false,font=normalsize,labelfont=sf,textfont=sf]{subfig}
\usepackage{textcomp}
\usepackage{stfloats}
\usepackage{url}
\usepackage{verbatim}
\usepackage{graphicx}
\usepackage{xcolor}
\usepackage{cite}
\usepackage{booktabs}
\usepackage[hidelinks]{hyperref}
\usepackage{makecell}
\begin{document}

\title{Spatial Speech Perception Systems: A Survey of Sound Source Localization, Directional Enhancement, and Speech Recognition}

\author{Pengyuan~Shao,
        Dimitrios~Kanoulas
        
\thanks{Pengyuan Shao and Dimitrios Kanoulas are with the Department of Computer Science, University College London, London, U.K.}
}

\markboth{
}{Shao \MakeLowercase{\textit{et al.}}: Spatial Speech Perception Pipelines}


\maketitle

\begin{abstract}
Robust speech understanding in real-world acoustic environments remains a fundamental challenge for intelligent auditory systems. Applications such as robot audition, hearing aids, assistive listening devices, teleconferencing systems, and voice-controlled assistants must operate in the presence of background noise, reverberation, competing speakers, and dynamic acoustic conditions to accommodate various real-world settings. Inspired by the human ability to selectively attend to relevant speakers in complex auditory scenes, increasing attention has been devoted to spatial speech perception, which exploits spatial information captured by microphone arrays to localize, enhance, and interpret target speech signals within complex acoustic scenes. Realizing this capability requires advances across multiple research areas, including sound source localization (SSL), speech enhancement and separation, and automatic speech recognition (ASR). Over the past decades, each of these fields has experienced substantial progress, driven by developments in array signal processing, deep learning, and large-scale speech modeling. As speech-enabled systems become increasingly autonomous and interactive, understanding the interplay among localization, enhancement, and recognition has become increasingly important for achieving robust speech perception in real-world environments.

Despite rapid progress in each area, the literature remains fragmented across multiple research communities, making it difficult to obtain a unified understanding of how these technologies contribute to robust speech perception systems. This paper presents a comprehensive survey of spatial speech perception systems, with particular emphasis on the roles of sound source localization, directional speech enhancement, and automatic speech recognition individually and within integrated processing frameworks. We review both classical signal-processing approaches and recent learning-based methods, covering microphone-array localization, beamforming and neural enhancement techniques, speech separation, and modern recognition architectures. Beyond component-level analysis, we examine robustness to noise and reverberation, multi-speaker operation, real-time constraints, computational efficiency, and the relationship between enhancement quality and downstream recognition performance. We further examine representative applications in robot audition, hearing assistance, smart speakers, and teleconferencing, highlighting how different system requirements influence the design of spatial speech perception pipelines. Finally, we identify key challenges and outline future research directions toward robust, low-latency, and perception-aware speech systems capable of reliable operation in complex acoustic environments.
\end{abstract}


\section{Introduction}

With the rapid development of speech and audio signal processing technologies, intelligent auditory systems are increasingly expected to operate beyond clean, close-talking, and single-speaker conditions. Modern speech-enabled systems, including hearing aids, assistive listening devices, teleconferencing platforms, smart speakers, wearable devices, and robots, must process speech in complex acoustic environments involving background noise, reverberation, interfering sources, and multiple competing speakers \cite{Cherry1953Cocktail,Bregman1990ASA,Bronkhorst2000Cocktail,HaebUmbach2021FarFieldASR}. Unlike conventional laboratory speech recognition settings, where the input signal is often assumed to be clean, well captured, and produced by a single speaker, real-world auditory systems must infer useful speech information from spatially distributed, degraded, and dynamically changing acoustic observations. Under these conditions, robust speech understanding requires not only recognizing what is being said, but also identifying where the relevant speech originates from and how it can be separated from other sound sources.

This problem is closely related to the long-standing challenge of listening in multi-talker acoustic scenes. Human listeners are often able to attend to a target speaker in the presence of competing voices and background noise by exploiting a combination of spectral, temporal, spatial, and contextual cues \cite{Cherry1953Cocktail,Bregman1990ASA,Bronkhorst2000Cocktail}. In machine listening systems, however, this capability remains difficult to reproduce, particularly when speech is captured at a distance or by compact microphone arrays with limited spatial resolution. Far-field speech signals are affected by room reverberation, source movement, sensor placement, device noise, and overlap between speakers \cite{HaebUmbach2021FarFieldASR,Barker2018CHiME5,Watanabe2020CHiME6}. As a result, the acoustic signal received by the system may contain a mixture of target speech, competing speech, diffuse noise, and reflected sound, making direct recognition unreliable.

Among the key approaches to real-world speech processing, spatial speech perception has attracted increasing attention as a framework for robust speech understanding in complex acoustic environments. Spatial speech perception exploits spatial information captured by microphone arrays to support the analysis, enhancement, and recognition of speech signals. In this setting, speech is not treated merely as a monaural waveform to be recognized independently, but as an acoustic event embedded in space and time. The spatial origin of a speaker, the propagation path between source and microphones, the interference structure of the scene, and the geometry of the microphone array all influence the quality and reliability of the captured signal \cite{BrandsteinSilvermanBook,Doclo2010HearingAidBeamforming,HaebUmbach2021FarFieldASR}. Therefore, robust speech understanding in realistic environments requires a processing perspective that connects acoustic sensing, spatial filtering, target speech extraction, and linguistic decoding.

Spatial speech perception is relevant to a wide range of speech and audio applications. In hearing aids and assistive listening devices, spatial selectivity can improve speech intelligibility by suppressing competing speakers and background noise while preserving useful binaural cues \cite{Doclo2010HearingAidBeamforming}. In teleconferencing and meeting transcription, multi-microphone processing can help separate overlapping speakers and improve recognition performance in distant conversational recordings \cite{Barker2018CHiME5,Watanabe2020CHiME6}. In smart speakers and voice-controlled assistants, spatial awareness can improve far-field speech capture under reverberant and noisy conditions \cite{HaebUmbach2021FarFieldASR}. In robot audition, spatial speech perception supports speaker localization, speech-driven navigation, human-robot interaction, and auditory scene understanding in dynamic environments, enabling robots to better interpret and respond to real-world acoustic events \cite{Nakadai2020RobotAuditionCASA}. Across these applications, the common objective is to transform degraded multichannel acoustic observations into reliable speech information under practical constraints such as latency, robustness, and limited computational resources.

\begin{figure*}
\centering
\includegraphics[width=0.99\linewidth]{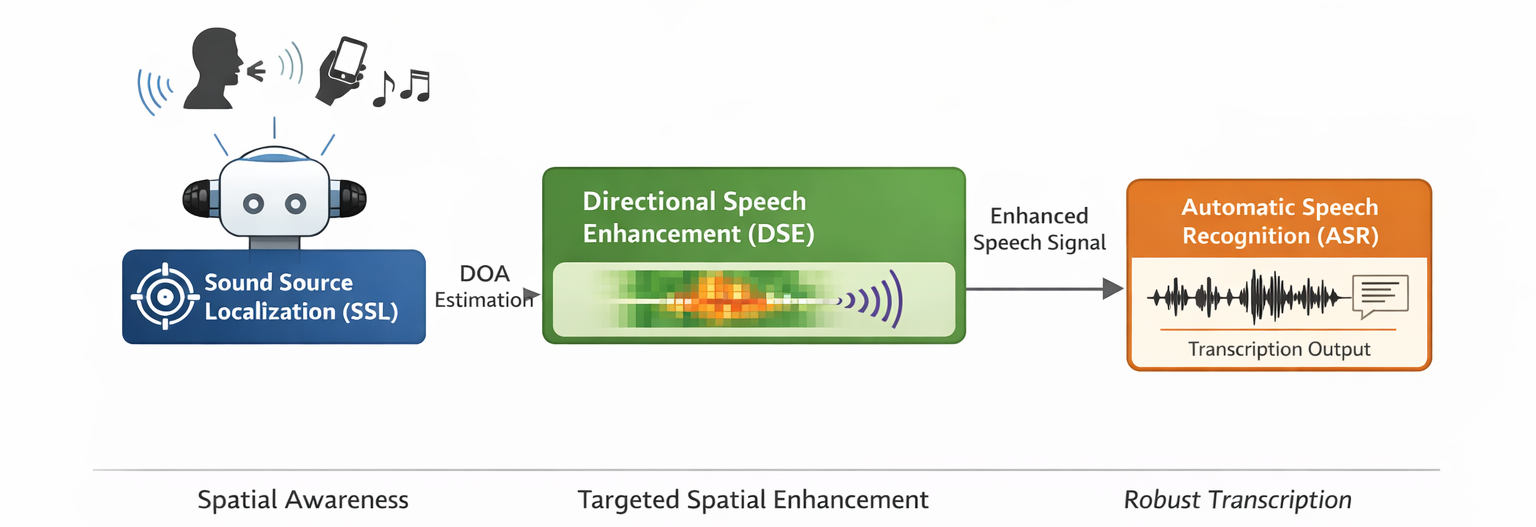}
\caption{Overview of a spatial speech perception pipeline, illustrating the integration of sound source localization (SSL), directional speech enhancement (DSE), and automatic speech recognition (ASR).}
\label{fig:pipeline}
\end{figure*}

This motivates the study of spatial speech perception systems, where acoustic signals are processed through a pipeline involving sound source localization (SSL), directional speech enhancement (DSE), and automatic speech recognition (ASR). As illustrated in Fig.~\ref{fig:pipeline}, SSL estimates the direction or position of acoustic sources from microphone array observations, DSE enhances the target speech signal by exploiting spatial information such as the direction of arrival (DOA), and ASR converts the resulting speech signal into text, commands, or other linguistic representations. This pipeline provides a structured view of how multichannel acoustic observations can be transformed into interpretable speech information.

\subsection{Background}

Traditional speech and audio processing systems have primarily focused on well-defined component-level objectives, such as speech quality improvement, acoustic source localization, speech enhancement, separation accuracy, or word-level recognition performance. In such systems, performance is generally measured using task-specific metrics, including signal-to-noise ratio (SNR), angular error, signal distortion ratio, perceptual quality, speech intelligibility, word error rate, throughput, or latency. However, modern intelligent auditory systems increasingly involve application-oriented objectives, where the value of the processed speech signal depends not only on waveform fidelity or component accuracy, but also on its usefulness for perception, interaction, reasoning, and downstream decision-making. For example, in speech-based interaction, a system may not need to reconstruct the exact acoustic waveform if it can reliably infer the speaker's intent. Conversely, high signal fidelity alone does not guarantee successful speech understanding if the recovered speech remains unintelligible, misrecognized, or attributed to the wrong speaker.

This shift introduces new requirements for spatial speech perception systems. Instead of treating speech as a passive single-channel signal to be enhanced or recognized independently, practical auditory systems must account for the complete information flow from acoustic capture to linguistic interpretation. The first challenge is spatial ambiguity. In realistic environments, multiple sound sources may be active simultaneously, and the target speaker may change over time. Without spatial awareness, a system may enhance or recognize the wrong source \cite{BrandsteinSilvermanBook,Grumiaux2022SurveyDeepSSL,Jekaterynczuk2024SSLSurvey}. The second challenge is acoustic degradation. Background noise, reverberation, and overlapping speech can significantly reduce the quality and intelligibility of the captured signal \cite{Deepclustering,DANet,DRN,CDUNet,MIDEANet}. The third challenge is recognition reliability. Even when a signal-level enhancement method improves perceptual quality, it may not necessarily reduce ASR errors or improve downstream task success \cite{Ahlawat2025ASRSurvey,pmlr-v202-radford23a}. The fourth challenge is real-time operation. Interactive speech systems require low and predictable latency, and long processing delays can reduce usability even when recognition accuracy is high \cite{yu2021fastemitlowlatencystreamingasr,su132212392}.

SSL, DSE, and ASR address different parts of this problem. SSL provides spatial sensing by estimating the source location or DOA from multichannel signals. Classical SSL methods, including time-difference-of-arrival estimation, generalized cross-correlation, beamforming-based localization, and subspace-based methods, rely on explicit acoustic models and microphone array geometry \cite{BrandsteinSilvermanBook,Knapp1976GCC,MUSIC_schmidt1986music}. These approaches often have interpretable structures and predictable computational properties, making them attractive for low-latency systems. However, their performance can degrade under strong reverberation, diffuse noise, or multiple simultaneous speakers. More recent learning-based SSL approaches attempt to learn spatial representations directly from multichannel audio features, improving flexibility in complex acoustic scenes, but often require large datasets and may introduce additional computational cost \cite{Grumiaux2022SurveyDeepSSL,Jekaterynczuk2024SSLSurvey}.

DSE forms the second stage of the spatial speech perception pipeline. Once the target direction is estimated or inferred, DSE aims to extract or enhance the speech signal corresponding to that direction. Classical beamforming methods such as minimum variance distortionless response (MVDR) and generalized sidelobe canceller (GSC) use spatial filtering to preserve the target direction while suppressing interference \cite{MVDR_1449208,Generalized_sidelobe_canceller}. These methods are mathematically grounded and computationally efficient, but they depend on accurate steering vectors, reliable covariance estimation, and appropriate array calibration. Neural enhancement methods, including masking-based, filtering-based, and DOA-conditioned architectures, have been introduced to improve robustness in challenging environments \cite{Deepclustering,DANet,FASNET,DRN,CDUNet,MIDEANet}. In particular, DOA-conditioned enhancement models explicitly use spatial information as input, making them highly relevant to integrated spatial speech perception systems.

ASR provides the linguistic decoding stage. Its role is to transform the enhanced speech signal into text, commands, or other linguistic representations. ASR has evolved from traditional statistical systems based on hidden Markov models and handcrafted acoustic features to modern end-to-end neural architectures based on CTC, RNN-T, attention mechanisms, Transformers, Conformers, and self-supervised speech representation learning \cite{rabiner1989tutorial,hinton2012deep,graves2006connectionist,graves2012sequencetransductionrecurrentneural,NIPS2017_3f5ee243,gulati2020conformerconvolutionaugmentedtransformerspeech,Baevski2020Wav2vec2}. These advances have significantly improved recognition performance under many conditions. Nevertheless, ASR remains vulnerable to noise, reverberation, domain mismatch, overlapping speech, and upstream enhancement artifacts \cite{Ahlawat2025ASRSurvey,pmlr-v202-radford23a}. This vulnerability is particularly important from a spatial speech perception perspective because the final objective is reliable speech understanding rather than signal reconstruction alone.

Although SSL, DSE, and ASR have each achieved substantial progress, practical auditory systems often require these technologies to operate together rather than as isolated modules. This creates a need to understand not only the performance of each component, but also how spatial estimates, enhanced speech signals, and recognition outputs interact across the complete processing chain. For example, an inaccurate or unstable DOA estimate may cause DSE to suppress the target speaker or enhance an interfering source. Similarly, an enhancement model that improves perceptual quality may still introduce artifacts that degrade ASR performance, while an ASR model that performs well on clean speech may be vulnerable to residual noise, reverberation, or separation errors. Therefore, spatial speech perception should be studied as an integrated system in which localization, enhancement, and recognition are jointly evaluated under robustness, latency, computational, and deployment constraints.

\subsection{Motivation}

The integration of spatial sensing, speech enhancement, and recognition is important for building robust speech perception systems in real-world acoustic environments. This integrated perspective not only improves target-speaker understanding under interference, but also highlights how localization, enhancement, and recognition should be jointly considered under recognition, latency, robustness, and deployment constraints.

First, spatial speech perception enables more reliable target-speaker understanding under interference. In many practical environments, the target speaker is not the only active source. Competing speakers, environmental sounds, device-generated noise, and reverberation can severely degrade both speech intelligibility and recognition accuracy. By estimating the spatial structure of the acoustic scene, SSL helps identify candidate target sources, while DSE uses this spatial information to improve the target-to-interference ratio before recognition \cite{Grumiaux2022SurveyDeepSSL,Jekaterynczuk2024SSLSurvey,DRN,CDUNet,MIDEANet}. This makes the recognition process more target-aware than applying ASR directly to raw microphone signals, particularly in far-field, multi-speaker, and noisy conditions.

Second, the pipeline supports recognition-oriented speech processing by prioritizing information that is useful for downstream understanding. In spatial speech perception systems, the objective is not necessarily to reconstruct a perfectly clean waveform, but to preserve the acoustic and linguistic cues required for accurate recognition and interaction. Directional enhancement can therefore be interpreted as a target-aware front-end that increases the relative prominence of the desired speech signal before ASR. This is important because signal-level improvements in perceptual quality do not always translate into lower word error rate or better task performance \cite{Ahlawat2025ASRSurvey,pmlr-v202-radford23a}. A pipeline-level perspective is therefore needed to understand how enhancement objectives should be aligned with recognition-oriented criteria.

Third, the pipeline exposes important system-level trade-offs that cannot be captured by evaluating each component in isolation. For example, an SSL module with high angular accuracy may still be unsuitable for interactive systems if it introduces excessive delay or unstable source estimates. A DSE model may improve perceptual speech quality but distort phonetic cues required by ASR. An ASR model may achieve high offline accuracy but be unsuitable for real-time interaction if it requires long input contexts or delayed decoding. Therefore, spatial speech perception should be evaluated using criteria that jointly consider spatial accuracy, enhancement quality, recognition accuracy, latency, robustness, and computational cost \cite{su132212392,yu2021fastemitlowlatencystreamingasr}.

Fourth, spatial speech perception provides a practical framework for deployable and resource-constrained auditory systems. In many real-world applications, acoustic sensing is performed on compact microphone arrays, wearable devices, smart speakers, hearing aids, or mobile robots, where computation, memory, energy consumption, and latency are limited. Considering SSL, DSE, and ASR as an integrated pipeline makes it possible to analyze where computation should be allocated, which stages should be performed locally, how much spatial information should be retained, and how enhancement and recognition models can be made efficient enough for real-time use. Although this survey primarily focuses on the algorithmic pipeline of SSL, DSE, and ASR, these deployment considerations motivate future research on efficient, adaptive, and perception-aware speech processing architectures for real-world operation.

\subsection{Contributions}

Through a systematic review and comparative analysis of the existing literature, this paper provides a unified knowledge framework for spatial speech perception systems, covering component technologies, system-level evaluation, integrated processing pipelines, applications, challenges, and future research directions. The main contributions are as follows.

\begin{itemize}
\item \textbf{Systematic review of component technologies:}
In Sections~\ref{sec2}, \ref{sec3}, and~\ref{sec4}, we review classical and learning-based methods for SSL, DSE, and ASR. For SSL, we discuss traditional approaches such as TDoA, GCC-PHAT, SRP-PHAT, MVDR-based localization, and subspace-based methods, together with learning-based CNN, CRNN, and attention-based models. For DSE, we examine filtering-based, masking-based, and DOA-conditioned enhancement approaches. For ASR, we summarize the evolution from traditional hybrid systems to CTC, RNN-T, Transformer, Conformer, self-supervised, and large-scale pretrained models.

\item \textbf{System-level evaluation for spatial speech perception:}
In Section~\ref{sec5}, we analyze existing works from the perspectives of robustness, real-time feasibility, noise tolerance, multi-speaker operation, and downstream recognition performance. In particular, we emphasize that spatial speech perception systems should be evaluated not only by component-level metrics, but also by system-level criteria such as latency accumulation, error propagation, recognition accuracy, target-speaker consistency, and recognition-oriented reliability. This perspective highlights the need to assess whether improvements in SSL or DSE actually translate into more robust speech understanding.

\item \textbf{Analysis of integrated spatial speech perception pipelines:}
In Section~\ref{sec6}, we survey representative systems that combine localization, enhancement, separation, and recognition. Rather than treating these components as independent modules, we analyze how spatial information is used to guide enhancement, how enhanced speech affects recognition, and how end-to-end or jointly optimized architectures attempt to align intermediate processing with downstream recognition objectives. This analysis clarifies the design trade-offs among interpretability, robustness, latency, computational efficiency, and recognition-oriented optimization.

\item \textbf{Applications, challenges, and future directions for robust auditory systems:}
In Section~\ref{sec7}, we discuss representative applications of spatial speech perception in hearing aids and assistive listening devices, robot audition, smart speakers and smart-home interfaces, teleconferencing and meeting transcription, auditory software interfaces, and wearable or mobile speech systems. We further identify key open challenges, including the mismatch between signal-level and task-level objectives, limited real-time evaluation, simplified acoustic assumptions, lack of standardized SSL--DSE--ASR pipeline benchmarks, robustness under noisy and dynamic multi-speaker conditions, multichannel microphone and array constraints, and resource limitations in embedded deployment. Finally, we outline future directions toward perception-aware spatial speech processing, including recognition-oriented enhancement, uncertainty-aware spatial sensing, multimodal scene-aware perception, joint localization--enhancement--recognition modeling, streaming and causal architectures, and robust low-latency speech understanding in complex real-world acoustic environments.

\end{itemize}

\section{Sound Source Localization}
\label{sec2}
\subsection{Traditional Methods}

From a spatial speech perception perspective, SSL provides spatial information that supports target-aware speech processing. The estimated direction of arrival (DOA), source position, or spatial probability map can be used to guide downstream directional speech enhancement and improve the reliability of speech recognition.

Sound source localization (SSL) is a fundamental spatial sensing task in spatial speech perception systems, aiming to estimate the direction or position of an acoustic source using signals captured by a microphone array. Traditional SSL methods are primarily rooted in array signal processing and explicit acoustic modeling, relying on assumptions about sound propagation, microphone geometry, and noise characteristics. These approaches have played a central role in practical speech and acoustic sensing systems due to their low computational requirements, deterministic structure, interpretability, and predictable latency \cite{BrandsteinSilvermanBook,Nakadai2020RobotAuditionCASA,7_8206494,51_grondin2022odasopenembeddedaudition}.

\subsubsection{Time-Difference-of-Arrival (TDoA) and Cross-Correlation Methods}

One of the earliest and most intuitive formulations of SSL is based on estimating the time difference of arrival (TDoA) between signals recorded by pairs of microphones. When a sound wave propagates across a microphone array, it reaches spatially separated microphones at slightly different times. These delays encode spatial information on the source direction relative to the array, and can be exploited to infer the direction of arrival (DOA) under the far-field assumption \cite{TDoA}.

Cross-correlation methods estimate TDoA by measuring the similarity between microphone signals as a function of time lag. The generalized cross-correlation (GCC) framework introduced frequency-domain weighting functions to improve robustness against noise and reverberation, with the phase transform (PHAT) weighting proposed to emphasize phase alignment while suppressing magnitude effects \cite{Knapp1976GCC}. The resulting GCC-PHAT method has become a canonical TDoA estimator in acoustic sensing due to its low computational cost, and has been widely adopted in real-time acoustic sensing, robot audition, and microphone-array speech processing systems.

Despite its popularity, GCC-PHAT exhibits several limitations. It typically assumes the presence of a dominant sound source, making it sensitive to interference from competing speakers. In addition, reverberation introduces multiple correlation peaks corresponding to reflections, which can lead to erroneous delay estimates. The discretization of TDoA values further limits angular resolution, especially for compact microphone arrays commonly mounted on mobile systems. Several correlation-based extensions of GCC-PHAT have been proposed to partially mitigate these issues, including modified spectral weightings and robust cross-correlation formulations that suppress reverberation-induced spurious peaks or reduce interference from competing speakers \cite{Knapp1976GCC,Rui2004TDE,Silverman1997RobustTDE}. However, these approaches remain fundamentally limited by their reliance on pairwise correlations and peak selection, motivating a transition toward beamforming-based methods that exploit array-wide spatial integration.

\subsubsection{Steered Response Power and Beamforming-Based Localization}

Beamforming-based SSL methods extend the TDoA concept by coherently combining signals from all microphones under hypothesized source directions. In this framework, the microphone array is virtually steered toward candidate directions, and a spatial response measure is evaluated to determine the most likely source location. Delay-and-sum beamforming represents the simplest instantiation of this idea and forms the basis of many classical spatial filtering approaches \cite{delay-and-sum}. Specifically, signals received at each microphone are time-aligned according to the expected propagation delays from a hypothesized direction and then summed, such that contributions from the true source add constructively while signals arriving from other directions are attenuated due to phase misalignment.

Beyond delay-and-sum, adaptive beamforming methods exploit second-order spatial statistics to enhance directional selectivity. A representative example is the Minimum Variance Distortionless Response (MVDR) beamformer \cite{MVDR_1449208}, which designs a spatial filter by minimizing the output power using the spatial covariance matrix, subject to a distortionless constraint on signals arriving from a hypothesized direction. MVDR, therefore, is capable of suppressing interference and background noise more effectively than fixed beamformers. MVDR has been successfully deployed in some robotic systems to suppress ego-noise and environmental interference, particularly in outdoor mobile robots \cite{8_lagacé2023egonoisereductionmobilerobot}. In practice, however, MVDR performance depends critically on accurate covariance estimation, which can be challenging in highly dynamic environments.

Building on the directional response of beamforming, sound source localization can be formulated by evaluating this response across candidate source directions. Steered Response Power (SRP) methods follow this idea by scanning the spatial domain and selecting the direction that maximizes the array response, directly producing a direction-of-arrival estimate. SRP-PHAT further combines this scanning strategy with PHAT-weighted cross-correlations, improving robustness in reverberant environments. Due to its ability to handle arbitrary array geometries and localize multiple speakers, SRP-PHAT has been widely used in real-time acoustic middleware such as ODAS, enabling real-time multi-source localization on mobile and embodied platforms \cite{51_grondin2022odasopenembeddedaudition}.

However, SRP-based methods typically rely on discretized spatial grids and precomputed TDoA tables, leading to substantial memory usage and computational overhead. To address these challenges, various acceleration strategies have been proposed, including hierarchical spatial search, coarse-to-fine resolution schemes, and lookup table optimization \cite{51_grondin2022odasopenembeddedaudition}. Despite these optimizations, SRP-based systems remain sensitive to strong noise and source overlap, which has motivated hybrid pipelines that combine beamforming with additional signal enhancement or learning-based components in real-world deployments \cite{6_liu2025soundsourcelocalizationhumanrobot}.

\begin{table*}[htbp]
\centering
\caption{Traditional signal-processing-based sound source localization methods in real-world spatial acoustic systems.}
\label{tab:traditional_ssl}
\resizebox{\textwidth}{!}{
\begin{tabular}{c c c c c c c c}
\toprule
\textbf{Paper} & \textbf{Year} & \textbf{Platform} & \textbf{Env.} & \textbf{Model} & \textbf{\# Mic} & \textbf{Max Src.} & \textbf{S/M} \\
\midrule
\cite{7_8206494}& 2017 & UAV & Outdoor & SEVD-MUSIC, iGSVD-MUSIC + ORPCA & 36 & 1 & Static \\
\cite{80_8967690} & 2019 & Humanoid & Indoor & DSVD-PHAT & 4 & 1 & Static \\
\cite{8_lagacé2023egonoisereductionmobilerobot}  & 2023 & UGV & Indoor & PCA + MVDR & 16 & 1 & Static \\
\cite{32_jiang2024duet-srp-phat} & 2024 & N/S   & Indoor & DUET + SRP-PHAT & 4  & 3 & Static \\
\cite{6_liu2025soundsourcelocalizationhumanrobot}  & 2025 & UGV & Outdoor & MVDR Beamformer + IRM & 16 & 1 & Static \\
\bottomrule
\end{tabular}
}
\end{table*}

\subsubsection{Subspace-Based Methods}
To achieve higher angular resolution and improved multi-source discrimination, subspace-based methods such as Multiple Signal Classification (MUSIC) were introduced. MUSIC exploits the eigenstructure of the spatial covariance matrix to separate signal and noise subspaces, leveraging the orthogonality between array steering vectors and the noise subspace to identify source directions \cite{MUSIC_schmidt1986music}. Originally developed for narrowband signals, MUSIC was later extended to broadband scenarios, including speech, through frequency-domain aggregation techniques \cite{WangKaveh1985Broadband}.

Subspace-based methods provide super-resolution capabilities under ideal conditions, enabling the separation of closely spaced sources beyond the limits of beamforming-based approaches. In robotics, MUSIC-based methods have been applied to platforms requiring high angular resolution under severe noise conditions, such as microphone-array-equipped UAVs for search and rescue tasks \cite{7_8206494}. However, classical MUSIC assumes spatially white noise, uncorrelated sources, accurate array calibration, and prior knowledge of the number of active sources. These assumptions are frequently violated in real-world acoustic environments, where noise is often spatially correlated and sound sources may overlap both temporally and spectrally.

To address these limitations, several extensions of MUSIC have been proposed. Generalized Eigenvalue Decomposition MUSIC (GEVD-MUSIC) introduces a noise reference covariance matrix and solves a generalized eigenvalue problem to better handle colored and spatially correlated noise \cite{GEVD_MUSIC_5354419}. It has been adopted in robotic audition systems to improve robustness under strong ego-noise, particularly on aerial robots where propeller noise dominates the acoustic scene \cite{7_8206494}.

Generalized Singular Value Decomposition MUSIC (GSVD-MUSIC) \cite{GSVD_6385494} further refines this approach by enforcing orthogonality between the signal and noise subspaces, leading to improved DOA estimation accuracy under challenging noise conditions. Incremental variants such as iGSVD-MUSIC have been proposed to enable adaptive subspace updates for streaming and time-varying scenarios, reducing computational overhead and improving responsiveness in dynamic acoustic scenes \cite{IGSVD_6942813}. 

Despite these advances, MUSIC-based methods remain computationally demanding due to the repeated eigenvalue decompositions required for subspace estimation, motivating the development of hybrid approaches that seek to reduce computational complexity through structural reuse or low-rank approximations.

\subsubsection{Hybrid Approaches}

Beyond the traditional categorization, several approaches have been proposed that explicitly integrate elements across paradigms, among which SVD-based formulations of SRP-PHAT provide a representative example. In conventional SRP-PHAT, pairwise GCC-PHAT responses are aggregated across microphone pairs and evaluated through a steered spatial response over a discretized spatial grid. While effective, this procedure relies heavily on dense spatial sampling and precomputed TDoA tables, resulting in substantial memory usage and computational overhead. To address these limitations, SVD-based formulations reinterpret the SRP-PHAT objective as a low-rank approximation problem, enabling the spatial response to be computed more efficiently through matrix factorization \cite{80_8967690}. By applying singular value decomposition to the phase-transformed cross-spectral matrices, these methods reduce the dimensionality of the localization problem while preserving the dominant spatial information relevant for DOA estimation.

In summary, Table \ref{tab:traditional_ssl} presents traditional signal-processing-based sound source localization methods applied in real-world systems, reporting for each work the publication year, system embodiment (if any), operating environment, core localization model, microphone array size, maximum number of simultaneously active sources, and whether the setup involves static or moving sources.

\subsection{Learning-based Methods}

Recent years have seen a growing shift from purely model-based sound source localization (SSL) toward learning-based approaches. Instead of relying on handcrafted spatial cues and analytical formulations, these methods leverage deep neural networks to learn discriminative spatial representations directly from data. In real-world spatial speech perception scenarios, this shift is motivated by the need to operate under conditions that violate classical assumptions, including strong reverberation, platform-induced noise, and multiple simultaneous sound sources.

\subsubsection{Convolutional Neural Networks (CNNs)}
Convolutional neural networks (CNNs) constitute one of the earliest deep learning architectures applied to sound source localization, primarily due to their ability to extract spatial patterns from structured time-frequency and inter-channel representations. An early and influential contribution is the work of Chakrabarty and Habets \cite{Chakrabarty_2017}, who proposed a CNN-based framework for broadband direction-of-arrival estimation using phase-based spatial features. Their method operates on phase maps constructed from the short-time Fourier transform of multichannel signals, capturing inter-microphone phase differences across frequency bands.

Building on this line of work, CNN-based SSL has been further explored in interactive acoustic settings. He et al. \cite{1_He_2018} proposed a deep neural network framework for multiple speaker detection and localization. Their system was evaluated using real-world recordings collected from a Pepper humanoid robot equipped with a four-microphone array. The CNN operates on sub-band cross-correlation features derived from microphone pairs, enabling the network to capture spatial delay patterns associated with multiple simultaneous speakers. In addition, a likelihood-based output encoding was introduced to allow localization of an arbitrary number of active sources, resulting in improved detection accuracy and reduced angular error compared to conventional methods.

CNN-based SSL has also been investigated for humanoid auditory perception in indoor environments. Boztas \cite{5_boztas2023sound} developed a learning-based localization system for a NAO robot using time-frequency features extracted via discrete wavelet transforms. These wavelet-based representations encode both spectral and temporal characteristics of the acoustic signals and are used as inputs to various neural architectures, including CNNs, LSTMs, and bidirectional LSTMs. Experiments conducted on real-world recordings demonstrated that learning-based models can achieve accurate spatial estimation in controlled indoor settings. However, evaluations were performed offline, and performance under outdoor or highly dynamic acoustic conditions was not explored.

In addition to direct DOA regression, CNNs have also been used as learned weighting modules that enhance classical correlation-based localization pipelines while preserving their structure and computational efficiency. Wang et al. \cite{33_9561885} proposed a speech-oriented extension of GCC-PHAT in which a lightweight CNN predicts frequency masks from magnitude spectra. These masks are applied to the normalized cross-power spectrum to emphasize speech-dominant regions before cross-correlation, improving the reliability of TDoA estimation under noise and reverberation. The method was evaluated on real-world recorded datasets and demonstrated improved localization accuracy compared to standard GCC-PHAT, while remaining compatible with real-time deployment.

\subsubsection{Convolution Recurrent Neural Networks (CRNNs)}

While CNNs are effective at extracting spatial features, they do not explicitly model temporal dependencies, which are critical for handling intermittent speech, overlapping sound events, and moving sources. To address this limitation, convolutional–recurrent neural networks (CRNNs) combine convolutional front-ends with recurrent layers such as LSTMs or GRUs, enabling temporal aggregation of spatial cues.

CRNN architectures form the backbone of the Sound Event Localization and Detection (SELD) framework, most prominently represented by SELDnet \cite{40_SELFnet}. SELDnet unified sound event detection and direction-of-arrival estimation within a single network and has served as an influential baseline architecture for DCASE Task 3. Through extensive evaluations on multi-source and reverberant datasets, SELDnet demonstrated that joint temporal–spatial modeling significantly improves localization stability compared to frame-wise methods. Recent extensions of CRNN-based SELD systems have further improved robustness through data generation and imbalance handling. Shin and Chun \cite{4_Shin2023SoundEL} proposed a residual CNN combined with temporal modeling and multi-generator data sampling to address the imbalance between real and synthetic recordings in DCASE datasets. 

Although SELDnet and its variants have been extensively evaluated in benchmark settings, their deployment on physical systems remains limited. The primary reasons include high computational cost and reliance on large labeled datasets.

\begin{table*}[htbp]
\centering
\caption{Learning-based sound source localization methods in real-world spatial acoustic systems.}
\label{tab:learning_ssl}
\resizebox{\textwidth}{!}{
\begin{tabular}{c c c c c c c c}
\toprule
\textbf{Paper} & \textbf{Year} & \textbf{Platform} & \textbf{Env.} & \textbf{Model} & \textbf{\# Mic} & \textbf{Max Src.} & \textbf{S/M} \\
\midrule
\cite{1_He_2018}  & 2018 & Humanoid & Indoor & CNN-GCCFB / TSNN-GCCFB & 4  & 2 & Static \\
\cite{40_SELFnet} & 2019 & N/S & Indoor & CRNN + ACCDOA / mACCDOA & 4 & 3 & Moving \\
\cite{33_9561885} & 2021 & Humanoid & Indoor & CNN Mask + GCC-PHAT-SM + MLP/SNN & 4 & 2 & Static \\
\cite{41_Du_NERCSLIP_task3_report} & 2022 & N/S & Indoor & ResNet--Conformer + ACCDOA Ensemble & 4 & 3 & Moving \\
\cite{31_10319788} & 2023 & N/S & Indoor & Mic-Pair Model + Array-Specific MLP & 2--4 & 2 & S+M \\
\cite{4_Shin2023SoundEL}  & 2023 & N/S & Indoor & RCNN + Transformer & 4  & 3 & Moving \\
\cite{5_boztas2023sound}  & 2023 & Humanoid & Indoor & CNN, LSTM, biLSTM, MLP & 4  & 1 & Moving \\
\cite{3_zhang2025multiple}  & 2024 & N/S & Indoor & CNN--Transformer & 18 & 3 & Static \\
\cite{2_akter2025hybrid}  & 2025 & N/S & Indoor & Hybrid CNN--LSTM, Patch Transformer & 1  & / & / \\
\cite{9_wang2025singlemicrophonebasedsoundsourcelocalization} & 2025 & Wheeled & Indoor & Filter-Attention CNN + EKF & 1  & 1 & Static \\
\cite{10_wang2025ipdnet2efficientimprovedinterchannel} & 2025 & N/S & Indoor & IPDnet2 & 5  & 1 & S+M \\
\cite{70_jazaeri2025multispeakerdoaestimationbinaural} & 2025 & N/S & Indoor & CRNN + Source Count Fusion & 3 & 2 & S+M \\
\cite{81_fu2025auralnethierarchicalattentionbased3d} & 2025 & N/S & Indoor & Hierarchical Attention Network (AuralNet) & 2 & 3 & S+M \\

\bottomrule
\end{tabular}
}
\end{table*}

\subsubsection{Attention-based SSL}
Attention mechanisms and transformer-based architectures have been introduced to overcome the limitations of recurrent models by enabling global modeling across time, frequency, and spatial channels. By adopting self-attention mechanisms, these approaches can capture long-range dependencies among time-frequency representations, facilitating more robust sound source localization in complex and dynamic acoustic environments.

Zhang et al. \cite{3_zhang2025multiple} proposed a CNN–Transformer framework for multiple sound source localization, combining sub-band spatial features derived from SRP-PHAT with multi-head self-attention to distinguish true spatial peaks from spurious reflections using sub-band spatial features. While the system achieved strong localization accuracy on simulated and real-world datasets, evaluations were conducted in wireless acoustic sensor network settings rather than on mobile or embedded platforms. The reliance on dense spatial grids and transformer-based inference introduces significant computational overhead, limiting real-time deployment on local platforms.

The DCASE SELD further popularized attention components for joint localization and detection. Shin and Chun \cite{4_Shin2023SoundEL} combine a convolutional front-end with a Transformer encoder to better capture global temporal context, and propose data-generation strategies to mitigate imbalance between real and synthetic data in SELD training. Similarly, Wang et al. employ a ResNet–Conformer backbone and strong augmentation and ensemble strategies \cite{41_Du_NERCSLIP_task3_report}. Such systems illustrate the strong performance of attention-heavy SELD architectures, but they are generally oriented toward SSL benchmark accuracy rather than the real-time constraints required by interactive spatial speech perception systems.

To conclude, attention-based methods demonstrate strong modeling capacity and improved performance in complex acoustic scenes, but practical acoustic systems often favor lighter architectures or hybrid designs that balance localization accuracy with real-time constraints. Table \ref{tab:learning_ssl} provides an overview of learning-based sound source localization approaches. Overall, SSL provides the spatial sensing foundation for spatial speech perception by estimating where acoustic sources are located. Traditional methods remain attractive for real-time and resource-constrained systems due to their deterministic structure and low computational cost, whereas learning-based methods offer stronger modeling capacity in complex acoustic scenes. However, localization alone does not directly improve the quality or intelligibility of the captured speech signal. In noisy and multi-speaker environments, the estimated spatial information must be further exploited to suppress interference and extract the target speech. This motivates the next stage of the spatial speech perception pipeline: directional speech enhancement, where SSL outputs such as DOA estimates are used to guide target-aware signal enhancement before downstream speech recognition.

\section{Directional Speech Enhancement}
\label{sec3}

Sound source localization (SSL) enables a spatial speech perception system to estimate the direction of arrival (DOA) of acoustic sources in its surrounding environment. While SSL provides crucial spatial awareness, it does not directly improve the quality of the recorded speech signal. In realistic acoustic environments, speech signals are often captured under challenging acoustic conditions involving background noise, reverberation, and multiple simultaneously speaking individuals. As a result, even when the direction of a target speaker is correctly estimated, the captured audio may still contain significant interference from competing sources. This degraded signal quality can negatively affect downstream automatic speech recognition (ASR) systems and reduce the reliability of downstream speech recognition.

To address this challenge, speech enhancement and speech separation techniques are commonly applied before ASR to improve the intelligibility and recognition reliability of the captured speech signal. Early deep learning approaches focused on the problem of multi-speaker speech separation, where the objective is to isolate individual speakers from a mixture without explicit knowledge of their spatial location. Methods such as Deep Clustering and Deep Attractor Networks learn discriminative representations of time–frequency (TF) components that allow overlapping speech signals to be separated. Although these techniques can effectively isolate different speakers, they do not explicitly exploit spatial information or directional cues.

In spatial speech perception systems equipped with microphone arrays, spatial information can be leveraged to further improve speech enhancement performance. When combined with SSL estimates, spatial processing techniques allow the system to selectively enhance speech originating from a particular direction while suppressing interference from other spatial locations. Consequently, many spatial speech processing pipelines integrate SSL with spatial filtering or neural enhancement models to perform directional speech enhancement before the enhanced signal is passed to ASR modules. Existing approaches for speech enhancement can be broadly categorized into three groups: spatial filtering approaches, time–frequency masking approaches, and DOA-conditioned directional enhancement methods.

\subsection{Filtering-based Approaches}

Filtering-based approaches perform directional enhancement by estimating spatial filters that operate directly on multichannel microphone signals. These methods originate from classical beamforming techniques and remain widely used in real-time spatial speech processing systems due to their strong theoretical foundations and low computational complexity.

One of the most well-known beamforming methods is the Minimum Variance Distortionless Response (MVDR) beamformer \cite{MVDR_1449208}. MVDR computes spatial filter weights that minimize the output signal power while maintaining a distortionless response for the target direction. By exploiting the spatial covariance structure of the microphone signals, MVDR suppresses interference and noise arriving from other directions while preserving the desired speech signal.

Another widely used approach is the Generalized Sidelobe Canceller (GSC) \cite{Generalized_sidelobe_canceller}, which reformulates the beamforming problem into a constrained optimization framework consisting of a fixed beamformer, a blocking matrix, and an adaptive interference cancellation stage. The upper path is a constrained beamformer that steers to the desired direction, producing a signal-plus-interference output. The lower path uses a blocking matrix orthogonal to the steering vector to suppress the desired signal and retain interference-dominant components, allowing the adaptive filter to focus solely on suppressing interference. A weight vector filters the noise in the lower path, which is subtracted from the upper path to remove interference.

With the rise of deep learning, neural beamforming approaches have been proposed to replace analytically derived spatial filters with data-driven models. A representative example is the Filter-and-Sum Network (FaSNet) \cite{FASNET}, which learns adaptive beamforming filters directly from multichannel input signals. FaSNet first selects one channel as the reference signal and computes cross-correlations with the other channels to extract spatial information about the sound sources. It then applies a neural network to estimate time-domain filters for reference channel. The similar procedure is then applied to other channels to generate corresponding filters. These filters are then applied to the input signals and summed to generate the enhanced speech output. By learning spatial filtering parameters from data, FaSNet can achieve strong performance while maintaining low latency.

Filtering-based approaches rely on different mechanisms for estimating spatial filters. Classical beamforming methods such as MVDR and GSC derive spatial filters analytically based on array geometry and steering vectors corresponding to the target direction. In contrast, neural beamforming approaches such as FaSNet learn adaptive spatial filters directly from multichannel data without explicitly computing steering vectors. While analytical methods offer strong theoretical guarantees and low computational complexity, data-driven models provide greater flexibility in complex acoustic environments where noise characteristics and spatial configurations may vary.

\subsection{Masking-based Approaches}

Masking-based approaches operate in the time-frequency (TF) domain and estimate a spectral mask that suppresses interference while preserving the target speech components. These methods typically apply the short-time Fourier transform (STFT) to the multichannel input signals and estimate masks for each time-frequency bin.

One influential masking-based method is Deep Clustering \cite{Deepclustering}, which addresses the problem of separating overlapping speech sources by learning discriminative embeddings for each TF bin. Instead of directly estimating masks, the network maps each TF bin of the mixture spectrogram into a high-dimensional embedding space. TF bins belonging to the same speaker are encouraged to have similar embeddings, while bins from different speakers are separated in the embedding space. During inference, clustering algorithms such as k-means are applied to the embeddings to group TF bins corresponding to different speakers, and binary masks are subsequently constructed from the clustering assignments. This approach allows the system to separate multiple speakers without requiring explicit knowledge of the target speaker or direction, and it has become a foundational method for deep learning–based speech separation.

Another important masking-based method is the Deep Attractor Network (DANet) \cite{DANet}, which extends the deep clustering framework by introducing attractor points in the embedding space. In DANet, the neural network learns embeddings for each TF bin similar to deep clustering, but instead of performing clustering during inference, a set of attractor points is estimated to represent the centers of different speakers in the embedding space. Each TF bin is then associated with the nearest attractor point, allowing masks to be generated directly through similarity measures. This approach eliminates the need for an external clustering algorithm during inference and improves computational efficiency. Deep attractor networks also enable end-to-end training and have demonstrated strong performance in multi-speaker speech separation tasks.

Recent work has extended TF masking methods to multichannel scenarios by incorporating spatial information into neural networks. One notable example is the Joint Spatial and Tempo-Spectral Nonlinear Filter (JNF) \cite{JNF}, which integrates spatial filtering and spectral enhancement into a unified neural network architecture. In contrast to traditional pipelines that separate beamforming and post-filtering stages, JNF jointly models spatial, temporal, and spectral relationships using recurrent neural networks. JNF processes time-frequency representations through wideband and narrowband branches and generates corresponding masks using recurrent and fully connected layers. These masks are then applied to the time-frequency representations to estimate the target speech and noise components. By jointly modeling spatial cues and spectral characteristics, masking-based methods can achieve improved enhancement performance compared with traditional beamforming techniques.

However, despite their strong enhancement performance, masking-based approaches present several limitations for real-time spatial speech perception systems. First, most TF masking models operate in the short-time Fourier transform (STFT) domain, which requires buffering audio frames to construct time–frequency representations. This frame-based processing introduces algorithmic latency that may affect responsiveness in interactive speech applications. Furthermore, many architectures rely on recurrent networks that exploit temporal context, which can further increase inference delay depending on whether causal or bidirectional models are used. Second, these approaches typically do not explicitly incorporate direction-of-arrival (DOA) information. Instead, spatial cues are implicitly learned from inter-channel time and phase differences during training. As a result, the system lacks explicit directional control and may struggle to selectively enhance arbitrary speakers in dynamic multi-speaker environments.

\subsection{DOA-conditioned Directional Enhancement}

Recent research has explored the explicit integration of directional cues into neural speech enhancement models. Unlike conventional masking or beamforming approaches that infer spatial information implicitly from inter-channel phase differences, time delays, or learned spatial representations, these methods directly incorporate direction-of-arrival (DOA) estimates as conditioning inputs to the neural network. Typically, the DOA angle obtained from an upstream sound source localization (SSL) module is encoded as a directional embedding and fused with acoustic features extracted from multichannel microphone signals. By explicitly providing spatial guidance to the model, these architectures allow the enhancement network to focus on signals arriving from a specified direction, making them well suited for spatial speech perception pipelines where localization and enhancement modules operate sequentially.

\begin{table*}[htbp]
\centering
\caption{Comparison of directional speech enhancement approaches for spatial speech perception.}
\label{tab:dse_comparison}
\begin{tabular}{lccccc}
\toprule
\textbf{Category} & \textbf{Methods} & \textbf{DOA} & \textbf{Real-time} & \textbf{Strength} & \textbf{Limitation} \\
\midrule
Filtering-based 
& MVDR, GSC, FaSNet 
& $\checkmark$ 
& High 
& \makecell{Low latency\\Interpretable} 
& Sensitive to DOA and noise estimation\\

Masking-based 
& DC, DANet, JNF 
& $\times$ 
& Low 
& Strong separation quality
& \makecell{High latency \\ No directional control} \\

DOA-conditioned 
& DRN, CDUNet 
& $\checkmark$ 
& Medium
& \makecell{Explicit spatial control\\ Improved target extraction}
& Depends on DOA accuracy\\

\bottomrule
\end{tabular}
\end{table*}

A representative DOA-conditioned architecture is the Directional Recurrent Network (DRN) \cite{DRN}, which incorporates explicit directional information into a neural speech enhancement model. In DRN, the multichannel microphone signals are first converted into frame-level acoustic features, which are then processed by a spatial filtering module to capture inter-channel spatial information. The target-speaker DOA is discretized and encoded as directional embeddings derived from azimuth and elevation angles. These embeddings are fused with the acoustic features within a recurrent neural network composed of stacked LSTM layers, allowing the model to integrate temporal speech patterns with directional cues. The network then predicts enhanced speech frames corresponding to the specified direction. By explicitly injecting DOA embeddings into the recurrent enhancement network, DRN demonstrates how directional information can guide neural models to focus on speech originating from a target direction.

Building on this idea of incorporating explicit directional cues into neural architectures, alternative designs have explored convolutional encoder–decoder networks to integrate spatial guidance. Another representative DOA-conditioned architecture is the Causal Directed U-Net (CDUNet) \cite{CDUNet}, which integrates directional beamforming cues with a convolutional encoder–decoder network. In CDUNet, the DOA estimated by the SSL module is used to generate steering vectors corresponding to the target direction. Instead of relying on a single steering vector, the model constructs three steering vectors representing the target direction and two neighboring directions around it. These three steering vectors approximate the spatial context around the target direction, enabling the model to capture signals arriving from a small angular neighborhood rather than a single direction. The spatial features derived from these steering vectors are combined with spectral features from the multichannel input and processed through a U-Net style convolutional encoder–decoder architecture. The encoder captures hierarchical spectral–spatial representations while the decoder reconstructs enhanced speech features corresponding to the target direction.

More recent work further extends DOA-conditioned enhancement by introducing additional directional conditioning signals beyond the DOA itself. One such approach introduces an end-to-end directional speech extraction model that conditions the enhancement network on both DOA and beamwidth embeddings \cite{MIDEANet}. In this architecture, the multichannel mixture is first processed to extract spectral features, while the target DOA and the desired spatial beamwidth are encoded through a clue encoder that converts directional parameters into learnable embeddings. These directional embeddings define a spatial region centered around the target DOA and are fused with acoustic features within the neural network. The model then processes the combined features through convolutional modules designed to capture both cross-frequency (crossband) interactions and local spectral dependencies (narrowband processing). By jointly modeling spatial cues and spectral structures, the network learns to extract speech signals originating within the specified directional region while suppressing interference from other directions. Unlike previous approaches that rely only on a single DOA cue or fixed neighboring directions, this adjustable beamwidth-aware conditioning explicitly defines the spatial extraction region around the target speaker, allowing the system to flexibly control the spatial focus of the enhancement process and improving robustness when multiple speakers are present near the target direction.

Table~\ref{tab:dse_comparison} summarizes the key characteristics of different directional speech enhancement approaches in terms of spatial modeling and real-time performance. In spatial speech perception pipelines, DSE acts as the bridge between spatial sensing and speech recognition: it transforms the spatial information estimated by SSL into an enhanced target speech signal that can be more reliably processed by downstream ASR systems. However, enhancement quality alone does not guarantee recognition accuracy, since signal-level improvements may not always translate into lower recognition error. This motivates the next stage of the pipeline, where ASR converts the enhanced acoustic signal into linguistic representations for downstream interpretation.

\section{Automatic Speech Recognition (ASR)}
\label{sec4}

Automatic Speech Recognition (ASR) is the linguistic decoding stage of spatial speech perception systems. While sound source localization provides spatial awareness and directional speech enhancement improves the quality of the target signal, ASR converts the enhanced acoustic signal into linguistic representations such as words, commands, or transcriptions. In spatial speech perception systems, this stage is critical because the final objective is not only to obtain an enhanced signal, but also to produce reliable linguistic outputs that support interaction, decision-making, and downstream applications.

In realistic acoustic environments, ASR systems must operate under challenging acoustic conditions, including background noise, reverberation, overlapping speakers, channel mismatch, and variable recording devices. These factors can significantly degrade recognition accuracy, especially when ASR is applied directly to raw microphone signals. Therefore, the effectiveness of ASR depends not only on the recognition model itself, but also on the quality and spatial consistency of the upstream SSL and DSE stages. From a pipeline perspective, ASR provides the final semantic output of the spatial speech perception pipeline, while its performance reflects the cumulative effects of localization accuracy, enhancement quality, and latency constraints.

Recent advances in deep learning have significantly transformed ASR, shifting from traditional hybrid systems toward end-to-end architectures that directly map acoustic signals to text. Approaches such as Connectionist Temporal Classification (CTC), Recurrent Neural Network Transducers (RNN-T), attention-based encoder--decoder models, Transformer-based architectures, and Conformer models have achieved substantial improvements in recognition performance. In particular, large-scale self-supervised and pretrained models, such as wav2vec 2.0, data2vec, and Whisper, have demonstrated strong generalization across diverse acoustic conditions, making them promising for real-world speech applications.

ASR systems can also be distinguished by their inference mode. Streaming ASR models process speech incrementally and emit partial or final recognition results as audio is received, making them suitable for interactive applications where low latency is required. Representative streaming architectures include traditional online GMM-HMM and DNN-HMM systems, as well as neural transducer models such as RNN-T and Conformer-T \cite{rabiner1989tutorial,hinton2012deep,graves2012sequencetransductionrecurrentneural,yu2021fastemitlowlatencystreamingasr}. Recent streaming adaptations of large-scale encoder-decoder ASR models have also been explored to reduce the latency of models that were originally designed for offline or utterance-level transcription, including Whisper-Streaming and Simul-Whisper \cite{machacek-etal-2023-turning,machacek-polak-2025-simultaneous}.
 In contrast, non-streaming ASR models typically process complete utterances or longer audio segments before producing a transcription. Attention-based encoder-decoder models, self-supervised models such as wav2vec 2.0 and data2vec, and large-scale models such as Whisper are commonly used in this mode \cite{NIPS2017_3f5ee243,Baevski2020Wav2vec2,pmlr-v162-baevski22a,pmlr-v202-radford23a}. This distinction is important for spatial speech perception systems because a model with high offline recognition accuracy may still be unsuitable for real-time interaction if it requires long input segments or delayed decoding. The real-time performance of both streaming and non-streaming ASR models is discussed in Section~\ref{sec5}.

\begin{figure*}
\centering
\includegraphics[width=0.95\linewidth]{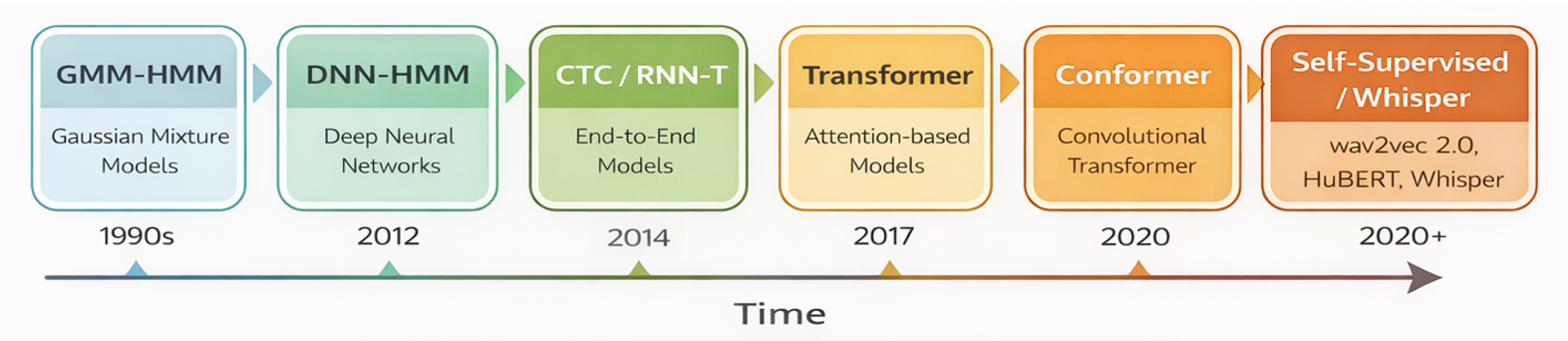}
\caption{Evolution of automatic speech recognition (ASR) architectures.}
\label{fig:asr_timeline}
\end{figure*}

In this section, we focus on the architectural structure of ASR models and review the evolution of ASR methodologies, covering both classical and modern approaches. This discussion provides a foundation for understanding how ASR can be integrated with upstream spatial sensing and enhancement modules in spatial speech perception pipelines. As illustrated in Fig.~\ref{fig:asr_timeline}, ASR architectures have progressively evolved from traditional hybrid GMM-HMM systems to modern end-to-end and self-supervised models.

\subsection{Traditional Hybrid ASR}

Early automatic speech recognition systems were predominantly based on hybrid statistical frameworks that combine acoustic modeling with probabilistic temporal sequence modeling. Among these, the Gaussian Mixture Model–Hidden Markov Model (GMM-HMM) architecture has been the most widely adopted and serves as the foundation of conventional ASR systems \cite{rabiner1989tutorial}. In this framework, raw speech signals are first transformed into compact acoustic representations, typically using handcrafted features such as Mel-Frequency Cepstral Coefficients (MFCCs), which are designed to capture perceptually relevant spectral characteristics of speech.

The Hidden Markov Model (HMM) is then used to represent the temporal structure of speech as a sequence of latent states corresponding to phonetic units. Transitions between these states are governed by state transition probabilities, enabling the model to capture the sequential nature of speech production. The observation likelihood of each acoustic feature vector given a hidden state is modeled using a Gaussian Mixture Model (GMM), which approximates the distribution of acoustic features associated with each phoneme. During decoding, the acoustic model is integrated with a pronunciation lexicon and a language model to determine the most probable word sequence.

To improve the modeling capacity of acoustic representations, deep neural networks were later introduced to replace GMMs, resulting in hybrid Deep Neural Network–Hidden Markov Model (DNN-HMM) systems \cite{hinton2012deep}. In this approach, deep architectures are used to model the complex, non-linear relationship between acoustic features and phonetic states, significantly improving recognition accuracy compared to traditional GMM-based systems. This transition marked an important step toward incorporating learning-based components into ASR pipelines.

However, both GMM-HMM and DNN-HMM frameworks fundamentally rely on handcrafted feature extraction, such as MFCCs, which may limit their ability to fully capture the rich structure of raw speech signals and adapt to diverse acoustic environments. This dependence on engineered features constrains the representational flexibility of the models and motivates the development of learning-based approaches that directly learn feature representations from data. These advances have led to the emergence of end-to-end ASR systems, which aim to jointly optimize acoustic modeling and sequence prediction within a unified framework.

\subsection{Learning-based ASR}
Learning-based ASR systems aim to directly map acoustic signals to text or token sequences by jointly learning feature representations and sequence prediction. Unlike traditional hybrid systems, these approaches replace handcrafted feature engineering and modular pipelines with neural network architectures that operate on raw or minimally processed audio.

\subsubsection{Connectionist Temporal Classification and RNN-Transducer ASR}
In early stage, learning-based ASR systems commonly adopt recurrent neural networks (RNNs) as the core architecture to model the temporal dynamics of speech signals. In these approaches, an encoder network, typically composed of stacked recurrent layers, processes input acoustic features and generates a sequence of hidden representations \cite{sak2014long}. These representations are then mapped to output tokens through different training objectives, among which Connectionist Temporal Classification (CTC) and the Recurrent Neural Network Transducer (RNN-T) are two widely used formulations.

In CTC-based systems, the RNN encoder produces frame-level predictions over output symbols (including a blank token), and the CTC loss function is used to compute the probability of all valid alignments between the input sequence and the target transcription \cite{graves2006connectionist}. This allows the model to be trained without explicit one-to-one frame-level alignment, significantly simplifying the training process. The CTC objective has been successfully applied in early end-to-end ASR systems such as Deep Speech \cite{hannun2014deepspeechscalingendtoend}, demonstrating that neural networks can directly learn speech-to-text mappings from data. However, the CTC formulation assumes that predictions at each time step are conditionally independent given the input sequence, and therefore does not explicitly model dependencies between output tokens.

To address this limitation, the RNN-Transducer (RNN-T) extends the RNN-based architecture by introducing an additional prediction network that conditions on previously generated output tokens \cite{graves2012sequencetransductionrecurrentneural, 8268935, pmlr-v48-amodei16}. The RNN-T model consists of three components: an encoder network that processes acoustic features, a prediction network that encodes output history, and a joint network that combines both representations to generate the next-token probability distribution. By incorporating both acoustic and label context, RNN-T enables joint modeling of input–output sequences within a unified framework. Furthermore, incorporating bidirectional recurrent architectures in the encoder allows the model to exploit both past and future acoustic context, which has been shown to significantly improve recognition performance in offline ASR settings \cite{graves2005framewise, 7472621}.

While RNN-based architectures effectively capture temporal dependencies in speech, their inherently sequential computation limits parallelization and scalability. In addition, their ability to model long-range dependencies is constrained, as information must be propagated step-by-step through the sequence, which can lead to degradation over long contexts. These limitations have motivated the development of attention-based and Transformer-based architectures, which replace recurrence with self-attention mechanisms to model global context more efficiently.

\subsubsection{Transformer-based ASR}

The Transformer architecture replaces recurrence with multi-head self-attention, allowing the model to directly capture global dependencies across the entire input sequence \cite{NIPS2017_3f5ee243}. In this framework, both the encoder and decoder are composed of stacked self-attention and feed-forward layers, enabling parallel computation over all time steps. In ASR, the Speech Transformer extends this architecture by applying the encoder–decoder framework to acoustic sequences, demonstrating that self-attention can effectively model speech signals without relying on recurrent structures \cite{8462506}. Compared to RNN-based models, Transformer ASR systems provide improved efficiency and stronger capability in modeling long-range temporal relationships.

Recent advances in ASR have been driven by self-supervised learning, where models are pretrained on large-scale unlabeled audio data and later fine-tuned for speech recognition tasks. These approaches enable the model to learn rich acoustic representations directly from raw waveforms, significantly reducing the need for annotated datasets. Representative models include wav2vec 2.0 and data2vec, which learn contextualized speech representations through pretraining objectives that capture temporal structure and contextual acoustic information \cite{Baevski2020Wav2vec2, pmlr-v162-baevski22a}. By leveraging large amounts of unlabeled data, self-supervised models improve robustness to noise, speaker variation, and domain shifts, making them highly effective for real-world speech perception applications.

Whisper represents a recent advancement in Transformer-based ASR, trained on a large-scale, diverse, and multilingual speech dataset \cite{pmlr-v202-radford23a}. The model adopts a sequence-to-sequence Transformer architecture and demonstrates strong generalization across different languages, accents, and acoustic conditions. Due to its large-scale training and robust representations, Whisper achieves high performance in noisy and real-world environments. Its ability to handle diverse inputs without task-specific tuning highlights the potential of large-scale pretrained models in advancing speech recognition systems.

\subsubsection{Conformer-based ASR}
The Conformer architecture extends Transformer-based ASR by integrating convolutional modules to better capture local temporal dependencies in speech signals \cite{gulati2020conformerconvolutionaugmentedtransformerspeech}. While self-attention effectively models global context, it lacks an inductive bias for local feature extraction, which is important for capturing phonetic and short-term spectral patterns. Conversely, convolutional neural networks (CNNs) are effective at modeling local structures but have limited ability to capture long-range dependencies due to their finite receptive fields, which require deeper architectures or larger kernels to expand, leading to increased computational cost. The Conformer addresses this limitation by combining multi-head self-attention with convolutional blocks within each layer, enabling the model to jointly capture both global and local dependencies.

By integrating self-attention with convolutional operations, the Conformer architecture highlights the complementary roles of global and local modeling in speech recognition. More generally, the progression from traditional hybrid systems to RNN-based architectures, and subsequently to attention and Transformer-based models, reflects a broader shift toward unified, data-driven approaches that jointly learn acoustic representations and sequence dependencies. These developments illustrate the increasing emphasis on capturing both long-range contextual information and fine-grained temporal patterns within a single framework, providing a foundation for modern end-to-end ASR systems.

However, despite these significant advances, not all ASR models are equally suitable for real-time spatial speech perception systems. More broadly, the suitability of SSL, DSE, and ASR cannot be fully assessed from component-level accuracy alone. A spatial speech perception system should be evaluated according to how reliably it transforms noisy multichannel acoustic observations into accurate linguistic outputs under practical constraints. This requires considering not only localization accuracy, enhancement quality, or word error rate in isolation, but also robustness to noise, real-time feasibility, latency accumulation, error propagation, and downstream recognition performance. Therefore, the next section presents a system-level evaluation perspective, summarizing evaluation metrics across individual components, reviewing existing evidence on noise robustness and real-time performance for SSL, DSE, and ASR, and discussing how these metrics relate to system-level recognition reliability.

\section{System-level Evaluation of Spatial Speech Pipelines}
\label{sec5}

A key challenge in spatial speech perception systems is that individual components are often evaluated using different task-specific metrics. Sound source localization (SSL) methods are commonly assessed using angular error, localization accuracy, source counting accuracy, or detection rate. Directional speech enhancement (DSE) methods are typically evaluated using signal-level or perceptual metrics such as signal-to-distortion ratio (SDR), scale-invariant signal-to-distortion ratio (SI-SDR), perceptual evaluation of speech quality (PESQ), short-time objective intelligibility (STOI), or enhancement latency. Automatic speech recognition (ASR) systems are primarily evaluated using word error rate (WER), character error rate (CER), real-time factor (RTF), partial recognition latency, or command success rate. While these metrics are useful for analyzing individual modules, they do not fully characterize the effectiveness of a complete spatial speech perception pipeline.

From a system-level spatial speech perception perspective, the final objective is to support reliable speech understanding under real-world acoustic and deployment constraints. Therefore, evaluation should consider how errors and delays accumulate across SSL, DSE, and ASR. For example, an inaccurate DOA estimate may guide DSE toward the wrong spatial region, causing target speech distortion or interference leakage, which may subsequently increase ASR error. Similarly, an enhancement model that improves signal-level metrics may not necessarily improve recognition accuracy if it distorts phonetic cues or introduces artifacts. These examples show that spatial speech perception systems require both component-level and system-level evaluation.

In this section, we summarize existing quantitative evidence on robustness and real-time feasibility, with particular emphasis on SSL and ASR, where reported results are more consistently available. For DSE, we discuss representative trends and evaluation metrics, while noting that directly comparable real-time and downstream ASR results remain less consistently reported across the literature. This motivates benchmark design that jointly evaluates spatial accuracy, enhancement quality, recognition performance, latency, and downstream recognition reliability.

\subsection{Evaluation Metrics Across SSL, DSE, and ASR}

Evaluation metrics for spatial speech perception systems can be broadly divided into three levels: component-level metrics, cross-stage metrics, and system-level pipeline metrics. Component-level metrics measure the performance of SSL, DSE, and ASR independently. Cross-stage metrics examine how the output of one stage affects the next stage. System-level metrics assess whether the complete pipeline successfully supports robust and low-latency semantic information exchange. This distinction is important because high performance at one stage does not necessarily guarantee high performance for the complete pipeline.

\begin{table*}[htbp]
\centering
\caption{Representative evaluation metrics for SSL, DSE, and ASR in spatial speech perception systems.}
\label{tab:evaluation_metrics}
\resizebox{\textwidth}{!}{
\begin{tabular}{p{2.2cm} p{3.6cm} p{4.8cm} p{5.2cm}}
\toprule
\textbf{Module} & \textbf{Common Metrics} & \textbf{What They Measure} & \textbf{Pipeline-Level Limitation} \\
\midrule
SSL 
& Angular error, localization accuracy, detection rate, source counting accuracy, tracking stability 
& Spatial accuracy, source detection, and temporal stability of DOA or position estimates 
& Accurate localization alone does not guarantee improved speech understanding; DOA errors may propagate to DSE and degrade ASR. \\[3.8em]

DSE 
& SDR, SI-SDR, PESQ, STOI, eSTOI, SIR, SAR, enhancement latency 
& Signal fidelity, perceptual quality, intelligibility, interference suppression, and processing delay 
& Signal-level improvement may not correspond to lower WER or better recognition reliability, especially when enhancement artifacts distort phonetic cues. \\[3.8em]

ASR 
& WER, CER, command success rate, intent accuracy, RTF, partial recognition latency 
& Transcription accuracy, linguistic/task accuracy, and real-time recognition feasibility 
& Recognition metrics often do not reveal whether errors originate from localization, enhancement, acoustic mismatch, or ASR model limitations. \\[3.8em]

End-to-end pipeline 
& Downstream WER after enhancement, total latency, robustness across SNRs, task success rate, task error rate 
& Overall ability to transform noisy acoustic input into reliable output under practical constraints 
& Requires integrated benchmarks and consistent reporting across SSL, DSE, and ASR, which remain limited in current literature. \\
\bottomrule
\end{tabular}
}
\end{table*}

For SSL, the most common metrics quantify the spatial accuracy and detection capability of the localization module. Angular error, often measured in degrees, evaluates the difference between the estimated and ground-truth DOA. Localization accuracy or recall measures whether the active source is correctly localized within a predefined angular tolerance. In multi-source scenarios, additional metrics such as source counting accuracy, detection rate, false alarm rate, and missed detection rate are used to evaluate whether the system can identify the correct number of active sound sources. For moving-source scenarios, tracking stability, temporal continuity, and source identity consistency may also be considered. These metrics are essential for assessing the spatial sensing capability of the system. However, from a pipeline perspective, SSL should also be evaluated according to whether its output is sufficiently accurate and stable to guide downstream DSE. A small localization error may be acceptable for scene awareness, but it may become harmful if the DSE module uses the DOA estimate to construct a narrow spatial filter.

For DSE, evaluation is usually based on signal-level quality, perceptual intelligibility, and enhancement efficiency. SDR and SI-SDR are commonly used to measure the degree of signal distortion or separation quality. PESQ and STOI provide perceptual estimates of speech quality and intelligibility. Other metrics, such as extended short-time objective intelligibility (eSTOI), signal-to-interference ratio (SIR), and signal-to-artifact ratio (SAR), may also be used depending on the task setting. These metrics are useful for measuring whether the enhanced signal is closer to the clean reference signal. Nevertheless, they are not always aligned with downstream semantic objectives. An enhancement model may improve SI-SDR or PESQ while introducing spectral artifacts that degrade ASR performance. Conversely, an enhancement output that sounds less natural may still preserve phonetic information and improve recognition. Therefore, DSE evaluation in a spatial speech perception pipeline should include not only signal-level metrics, but also downstream ASR performance and latency.

For ASR, the most widely used metrics are WER and CER, which measure transcription errors at the word and character levels. These metrics directly reflect the transcription accuracy of the system. In command-based or task-oriented jobs, command success rate, intent recognition accuracy, or semantic error rate may be more informative than WER alone, because the final objective is often to recover the intended meaning rather than produce a perfect transcription. In addition, real-time ASR systems are commonly evaluated using RTF, streaming latency, and partial recognition latency. These metrics are particularly important for interactive speech systems, where delayed recognition can reduce system responsiveness even if the final transcription is accurate.

A key limitation of conventional evaluation practice is that SSL, DSE, and ASR are often assessed using separate datasets, different acoustic conditions, and different assumptions. SSL studies may focus on angular accuracy under controlled source positions, DSE studies may evaluate enhancement quality using clean reference signals, and ASR studies may report WER on speech recognition benchmarks that do not include spatial localization or directional enhancement. As a result, it is difficult to determine how improvements in one module affect the complete pipeline. This fragmentation is problematic for spatial speech perception systems, where the final performance depends not only on the accuracy of individual components, but also on the interaction among spatial sensing, enhancement, and semantic decoding.

A system-level evaluation framework should therefore include both component-level and pipeline-level metrics. At the component level, SSL should be evaluated using spatial accuracy, source detection, and tracking stability; DSE should be evaluated using enhancement quality, intelligibility, and processing delay; and ASR should be evaluated using recognition accuracy, latency, and streaming capability. At the pipeline level, however, the system should be evaluated using end-to-end recognition accuracy, total processing latency, robustness under noise and reverberation, and recognition reliability and task success. In particular, downstream WER, CER, intent accuracy, or command success rate after enhancement provides a more direct measure of whether DSE actually improves downstream recognition performance. Similarly, evaluating recognition performance under different DOA errors can reveal how sensitive the full pipeline is to localization uncertainty.

Recent integrated pipelines further illustrate the importance of system-level evaluation (Detailed architectures of recent integrated pipelines are in Section \ref{sec6}). Instead of optimizing SSL, DSE, and ASR independently, some systems introduce joint training objectives that combine localization, separation, enhancement, and recognition-related losses, while others optimize the front-end processing chain directly according to downstream ASR performance. For example, a model may jointly predict DOA and separated speech \cite{MSDET}, use DOA estimates to guide spatial filtering \cite{9414187, 11329501}, or train the localization–enhancement–recognition chain using recognition-oriented objectives such as CTC loss or WER-related criteria \cite{9414243}. These designs indicate that the final goal of the pipeline is not merely accurate localization or high signal fidelity, but reliable speech recognition under practical constraints. Therefore, pipeline-level evaluation should examine whether improvements in intermediate representations lead to lower recognition error, higher task success rate, and more robust speech understanding under noise, interference, and latency constraints.

Table~\ref{tab:evaluation_metrics} summarizes representative metrics across SSL, DSE, and ASR, together with their relevance to spatial speech perception systems. The table highlights that each component has its own mature evaluation criteria, but a complete system-level assessment requires linking these metrics across stages. In particular, component-level metrics should be interpreted according to their influence on downstream processing: SSL accuracy should be analyzed in terms of its effect on enhancement, DSE quality should be analyzed in terms of its effect on recognition, and ASR accuracy should be analyzed together with latency, robustness, and semantic reliability.

Overall, the evaluation of spatial speech perception systems should shift from isolated component assessment toward end-to-end recognition reliability. This does not mean that component-level metrics are unnecessary; rather, they should be embedded within a pipeline-level evaluation framework. Such a perspective is essential for designing spatial speech perception systems that are not only accurate in individual tasks, but also efficient and reliable for downstream applications in realistic acoustic environments. Based on this evaluation perspective, the following subsections examine two key practical dimensions: real-time feasibility and noise robustness. We focus primarily on SSL and ASR, where quantitative results are more consistently reported in the literature, while discussing DSE in relation to its role as the intermediate enhancement stage.

\subsection{Evaluation of Real-Time Feasibility and Noise Robustness}

Real-time feasibility and noise robustness are two central requirements for spatial speech perception systems. In interactive speech scenarios, the system must process acoustic input with sufficiently low latency so that speech interaction remains responsive. At the same time, the system must remain robust under adverse acoustic conditions such as background noise, reverberation, overlapping speakers, and platform-induced noise. These requirements are particularly important because spatial speech perception is not only concerned with component-level accuracy, but also with the reliable transformation of noisy acoustic observations into semantic information.

In the following, we summarize reported evidence for SSL, DSE, and ASR from the perspectives of latency and noise tolerance. For SSL and ASR, quantitative results are more consistently available in the literature. For DSE, direct comparison is more difficult because studies often use different datasets, signal-level metrics, enhancement objectives, and hardware settings. Therefore, the DSE subsection focuses on representative evaluation trends and explains why enhancement should be assessed not only by signal fidelity, but also by downstream recognition performance.

\subsubsection{SSL}

While Sections~\ref{sec2} reviewed SSL methods from algorithmic and modeling perspectives, real-world spatial speech perception systems introduce additional constraints that fundamentally shape which approaches are practical. In particular, SSL must provide spatial estimates with sufficiently low latency and adequate robustness to adverse acoustic conditions. This is because DOA estimates are often used as spatial side information for downstream directional speech enhancement. If SSL is slow, unstable, or inaccurate under noise, the error may propagate to DSE and subsequently degrade ASR performance.

\begin{table*}[htbp]
\centering
\caption{Sound source localization systems with reported real-time performance and noise tolerance.}
\label{tab:ssl_combined}
\begin{tabular}{c c c c c c}
\toprule
\textbf{Paper} &
\textbf{Year} &
\textbf{Platform} &
\textbf{Method Type} &
\textbf{Real-time Latency} &
\textbf{Noise Tolerance (SNR dB)} \\
\midrule

\cite{7_8206494} &
2017 &
UAV &
Traditional &
\begin{tabular}[c]{@{}c@{}}
SEVD: $<1$ s \\
iGSVD: $2\sim3$ s
\end{tabular} &
\begin{tabular}[c]{@{}c@{}}
SEVD: $\geq 0$ dB \\
iGSVD: $\geq -20$ dB
\end{tabular}
\\[3.0mm]

\cite{80_8967690} &
2019 &
N/S &
Traditional &
\begin{tabular}[c]{@{}c@{}}
GSVD-MUSIC: $23.3$ ms\\
DSVD-PHAT: $0.093$ ms
\end{tabular} &
\begin{tabular}[c]{@{}c@{}}
Evaluated at $-10\sim20$ dB\\
GSVD-MUSIC: high-noise setting\\
DSVD-PHAT: low-noise setting
\end{tabular}\\[4.0mm]

\cite{8_lagacé2023egonoisereductionmobilerobot} &
2023 &
UGV &
Traditional &
0.2 s per 0.5 s audio &
Evaluated at $-5$ to $-10$ dB
\\[1.5mm]

\cite{6_liu2025soundsourcelocalizationhumanrobot} &
2025 &
UGV &
Traditional &
$2\sim3$ s per command &
$\geq 1$ dB
\\

\midrule

\cite{33_9561885} &
2021 &
Humanoid &
Learning &
Real-time claimed, exact latency not reported &
$\geq 0$ dB
\\

\cite{31_10319788} &
2023 &
ReSpeaker &
Learning &
$\sim 0.2$ s output rate  &
$\geq 10$ dB
\\

\cite{70_jazaeri2025multispeakerdoaestimationbinaural} &
2024 &
N/S &
Learning &
Offline, exact latency not reported &
Evaluated at $5\sim15$ dB
\\

\cite{9_wang2025singlemicrophonebasedsoundsourcelocalization} &
2025 &
Wheeled &
Learning &
Real-time claimed, exact latency not reported &
N/A
\\

\cite{10_wang2025ipdnet2efficientimprovedinterchannel} &
2025 &
N/S &
Learning &
$\sim$0.1 s output rate &
Evaluated at $-5\sim15$ dB
\\

\cite{81_fu2025auralnethierarchicalattentionbased3d} &
2025 &
N/S &
Learning &
Offline, exact latency not reported &
Evaluated at $0\sim20$ dB
\\

\bottomrule
\end{tabular}
\end{table*}

As summarized in Table~\ref{tab:ssl_combined}, traditional signal-processing-based SSL methods generally provide more explicit latency reporting and often more predictable real-time behavior. Several correlation-based and subspace-based approaches achieve millisecond-level processing times, while other systems report bounded processing times per command or per audio segment. These latency values are usually easier to interpret because traditional SSL pipelines have deterministic computational structures. In contrast, many learning-based methods report approximate output rates, provide limited timing breakdowns, or are evaluated offline. Although several neural SSL systems claim real-time feasibility, their latency reporting is often less standardized. This comparison suggests that traditional SSL methods currently remain attractive for latency-sensitive spatial speech perception, especially when predictable runtime is required.

Regarding noise robustness, Table~\ref{tab:ssl_combined} indicates that traditional approaches have often been evaluated under more severe noise conditions, including negative SNR regimes. Phase-transform weighting, beamforming, and subspace-based techniques are frequently tested in adverse acoustic scenarios, demonstrating their ability to operate under strong background interference. By comparison, learning-based methods are more commonly evaluated within moderate SNR ranges, typically from 0~dB to 20~dB, with fewer reported results under strongly negative SNR conditions. This does not necessarily mean that learning-based SSL is less robust; rather, it reflects differences in evaluation practice and dataset design. Overall, the available evidence suggests a practical tendency: traditional SSL methods emphasize computational efficiency and robustness under low-SNR conditions, whereas learning-based approaches provide stronger modeling capacity but often require more careful reporting of latency, robustness, and generalization. For a spatial speech perception pipeline, these metrics, together with DOA accuracy, should be considered jointly because they directly affect the subsequent enhancement and recognition stages.

\subsubsection{DSE}

Compared with SSL and ASR, real-time feasibility and noise robustness are more difficult to compare consistently for directional speech enhancement. This is because DSE studies often differ substantially in microphone configuration, input signal representation, acoustic scene complexity, target definition, and evaluation protocol. Some works evaluate enhancement using clean reference signals and report signal-level metrics such as SDR, SI-SDR, PESQ, STOI, or eSTOI, whereas others focus on perceptual quality, separation performance, or downstream ASR improvement. In addition, latency reporting is often incomplete, especially for neural enhancement models that operate in the time--frequency domain or use non-causal temporal context.

From the perspective of real-time feasibility, filtering-based approaches such as MVDR and GSC remain attractive because they are mathematically structured and computationally efficient. Once the steering vector or target direction is available, these methods can perform spatial filtering with relatively predictable latency. This makes them suitable for interactive speech systems where enhancement must be performed before ASR without introducing excessive delay. However, their performance depends strongly on accurate DOA estimation, reliable covariance estimation, and stable microphone array calibration. In dynamic acoustic scenes, errors in these quantities may reduce interference suppression or distort the target speech signal.

Masking-based neural enhancement methods generally provide stronger separation and enhancement capability, especially in multi-speaker and noisy environments. However, they often operate on STFT representations and may require temporal context to estimate reliable masks. This introduces algorithmic latency due to frame buffering, windowing, and overlap-add reconstruction. If bidirectional recurrent layers or non-causal attention mechanisms are used, additional future context may be required, making such systems less suitable for low-latency speech interaction. Therefore, although masking-based methods may improve signal-level metrics, their real-time feasibility depends on whether the architecture is causal, whether the frame size and hop size are small enough, and whether inference can be performed efficiently on the target hardware.

DOA-conditioned enhancement methods provide a promising compromise for spatial speech perception systems. By explicitly conditioning the enhancement model on the target DOA, these methods can use spatial information from SSL to guide target-aware extraction. This offers greater directional control than generic masking-based separation and greater flexibility than purely analytical beamforming. However, their performance is inherently coupled to SSL accuracy. If the estimated DOA is inaccurate or unstable, the enhancement model may focus on the wrong direction or suppress the target speech. Therefore, DSE robustness should be evaluated not only under different SNRs, but also under controlled DOA perturbations that simulate upstream localization errors. As for latency, recent work \cite{MIDEANet} has designed architectures that reduce model size and FLOPs through frequency-time pooling to improve inference efficiency.

Overall, DSE acts as the intermediate stage between spatial sensing and speech recognition. Its role is to convert DOA or spatial information into an enhanced target speech signal that supports reliable ASR. Future DSE evaluations should therefore report not only enhancement quality, but also latency, causal or non-causal processing mode, robustness across SNR levels, sensitivity to DOA errors, and downstream ASR performance. Such reporting would make it possible to determine whether a DSE method truly improves the reliability of spatial speech perception rather than only improving isolated signal-level metrics.

\subsubsection{ASR}

Real-time performance is a fundamental requirement for spatial speech perception systems, since delayed recognition can reduce interaction quality and responsiveness. A commonly used metric for computational efficiency is the real-time factor (RTF), defined as the ratio between processing time and input audio duration. An RTF value below 1 indicates that the system can process speech faster than real time, which is generally desirable for interactive speech systems.

For streaming ASR systems, a more interaction-oriented measure is partial recognition (PR) latency, which quantifies the time difference between the end of a spoken utterance and the moment when the last token is emitted in the finalized recognition result. Unlike throughput-based metrics, PR latency better reflects the delay perceived by users during real-time speech interaction. Therefore, both RTF and PR latency are important for evaluating whether ASR models can operate as the semantic decoding stage of a real-time spatial speech perception pipeline.

\begin{table*}[htbp]
\centering
\caption{Real-time feasibility of representative ASR models.}
\label{tab:asr_latency}
\resizebox{\textwidth}{!}{
\begin{tabular}{l l l l l}
\toprule
\textbf{Model} & \textbf{Mode} & \textbf{Hardware} & \textbf{Reported Metric} & \textbf{Reported Value} \\
\midrule
GMM-HMM (Kaldi) \cite{9291818}
& Streaming
& CPU
& RTF / processing time
& RTF $\approx 1.0$; $\sim$1000 ms per 1 s audio \\

DNN-HMM \cite{wong2024syllablebaseddnnhmmcantonese}
& Streaming
& CPU
& RTF / processing time
& RTF $\approx 0.8$--$1.4$; $\sim$800--1400 ms per 1 s audio \\

\midrule
RNN-T \cite{yu2021fastemitlowlatencystreamingasr}
& Streaming
& GPU/CPU
& PR50 latency
& $\sim$190 ms \\

Transformer-T \cite{yu2021fastemitlowlatencystreamingasr}
& Streaming
& GPU
& PR50 latency
& $\sim$220 ms \\

Conformer-T \cite{yu2021fastemitlowlatencystreamingasr}
& Streaming
& GPU
& PR50 latency
& $\sim$150 ms \\

\midrule
wav2vec 2.0 \cite{Andrew_2025}
& Non-streaming
& A5000 GPU
& Throughput / processing time
& $\sim$3 ms per 1 s audio \\

wav2vec 2.0 \cite{Andrew_2025}
& Non-streaming
& 2080Ti GPU
& Throughput / processing time
& $\sim$4--5 ms per 1 s audio \\

\midrule
Whisper Large-v2 \cite{faster-whisper}
& Non-streaming
& GPU (FP16)
& Processing time
& $\sim$183 ms per 1 s audio \\

Whisper Large (faster) \cite{faster-whisper}
& Non-streaming
& GPU (INT8)
& Processing time
& $\sim$76 ms per 1 s audio \\

Whisper Small \cite{faster-whisper}
& Non-streaming
& CPU
& Processing time
& $\sim$536 ms per 1 s audio \\

Whisper.cpp \cite{faster-whisper}
& Non-streaming
& CPU optimized
& Processing time
& $\sim$135--165 ms per 1 s audio \\
\bottomrule
\end{tabular}
}
\end{table*}

As shown in Table~\ref{tab:asr_latency}, traditional hybrid systems such as GMM-HMM and DNN-HMM can operate close to real time on CPU platforms, with latency on the order of 800--1400~ms per second of audio. Although these systems can support streaming decoding, their latency is often influenced by decoding complexity, search graphs, and language models. In contrast, modern end-to-end streaming architectures such as RNN-T, Transformer-T, and Conformer-T achieve substantially lower partial recognition latency, with reported values around 150--220~ms. These improvements are largely attributed to neural architectures that support incremental decoding and reduce reliance on complex modular decoding pipelines.

For non-streaming models, latency characteristics differ significantly depending on hardware and evaluation methodology. Self-supervised models such as wav2vec~2.0 can achieve very low processing times under GPU batch inference, but such measurements often reflect throughput-oriented evaluation rather than user-perceived streaming latency. Similarly, large-scale Transformer models such as Whisper demonstrate competitive latency under optimized GPU inference, but they are primarily designed for offline or utterance-level processing and may require access to complete speech segments. For real-time spatial speech perception systems, streaming capability and predictable low-latency behavior are therefore more important than throughput alone.

From a system-level perspective, models such as RNN-T and Conformer-T are particularly suitable for interactive scenarios because they support incremental inference and maintain latency well below real-time thresholds. In contrast, non-streaming models may achieve high recognition accuracy but require additional deployment strategies, such as chunked inference, voice activity detection, hardware acceleration, or model compression. Overall, achieving $RTF<1$ is a necessary but not sufficient condition; practical ASR systems should also provide stable streaming latency, robustness to enhanced speech artifacts, and compatibility with upstream SSL and DSE modules.

\begin{table*}[htbp]
\centering
\caption{Noise robustness comparison of representative ASR models.}
\label{tab:asr_noise}
\begin{tabular}{l l l}
\toprule
\textbf{Model} & \textbf{Reported Robustness Range}  & \textbf{Key Observation} \\
\midrule
Conformer-1 \cite{zhang2024conformer1robustasrlargescale} & $>$ 0 dB & Stable performance across moderate SNR levels \\
wav2vec 2.0 \cite{wang2022wav2vecswitchcontrastivelearningoriginalnoisy} & $>$ 5 dB & Evaluated mainly under moderate noise conditions \\
Whisper base.en \cite{simic2023selfsupervisedadaptiveavfusion} &  $>$ -5 dB & Performance degrades noticeably as SNR decreases \\
Whisper small.en \cite{simic2023selfsupervisedadaptiveavfusion} &  $>$ -10 dB & Improved tolerance to lower SNR compared to base variant \\
Whisper + StoRM \cite{10974078} &  $>$ -10 dB & Robustness extends toward lower SNR conditions \\
\bottomrule
\end{tabular}
\end{table*}

Due to the lack of standardized reporting, comprehensive noise robustness results are not available for all ASR models. Table~\ref{tab:asr_noise} therefore summarizes representative results collected from the literature under varying SNR conditions. The available evidence shows that many ASR models operate reliably only above moderate SNR levels, typically above 5--10~dB, with performance degrading as noise increases. Whisper-based models exhibit stronger robustness as SNR decreases, especially near and below 0~dB. Increasing model size can improve robustness, while additional enhancement or adaptation methods such as StoRM can further extend the effective operating range toward lower SNR conditions.

These findings suggest that, despite strong performance in clean or moderately noisy environments, current ASR systems remain vulnerable under highly adverse acoustic conditions. This reinforces the importance of upstream SSL and DSE modules in spatial speech perception systems.

\subsection{Lessons Learned}

The evaluation results discussed above reveal that spatial speech perception systems cannot be adequately assessed through isolated component-level metrics alone. Although SSL, DSE, and ASR each have mature evaluation criteria, their practical value depends on how well they support the complete information flow from acoustic sensing to semantic decoding. Several lessons can be drawn from the comparison of real-time feasibility, noise robustness, and downstream recognition reliability.

\begin{itemize}
    \item \textbf{Component-level accuracy is necessary but insufficient.}
    SSL accuracy, DSE quality, and ASR recognition performance are all important, but none of them alone determines the effectiveness of the complete spatial speech perception pipeline. A localization method with low angular error may still be unsuitable if its latency is high or if its DOA estimates are unstable under noise. Similarly, an enhancement method with improved SI-SDR or PESQ may not necessarily reduce WER if it introduces artifacts that distort phonetic information. Therefore, component-level metrics should be interpreted according to their impact on downstream stages.

    \item \textbf{Real-time feasibility remains unevenly reported across modules.}
    Traditional SSL methods often provide more explicit and predictable latency reporting due to their deterministic signal-processing structure. In contrast, many learning-based SSL and DSE approaches either report approximate output rates or are evaluated offline. ASR studies provide more standardized latency metrics, such as RTF and partial recognition latency, but comparisons between streaming and non-streaming models remain difficult. For interactive speech systems, reporting only throughput is insufficient; latency should be measured in a way that reflects user-perceived responsiveness.

    \item \textbf{Noise robustness should be evaluated across the complete pipeline.}
    Existing SSL and ASR studies often report robustness under different SNR ranges, but these evaluations are usually conducted independently. In a spatial speech perception pipeline, noise affects not only each module separately but also their interactions. Noise may degrade DOA estimation, inaccurate DOA may degrade enhancement, and residual interference or enhancement artifacts may increase ASR errors. Therefore, robustness should be evaluated under consistent acoustic conditions across SSL, DSE, and ASR.

    \item \textbf{Streaming capability is critical for interactive deployment.}
    ASR models with strong offline accuracy may not be suitable for real-time speech interaction if they require complete utterances or large computational resources. Streaming models such as RNN-T and Conformer-T are more aligned with interactive jobs because they support incremental decoding and lower partial recognition latency. Similarly, SSL and DSE modules should be designed with causal or low-latency processing when the target application requires real-time interaction.

    \item \textbf{Semantic reliability should become a central evaluation objective.}
    The final goal of spatial speech perceptions is not only to localize a source, enhance a waveform, or transcribe speech, but to reliably recover task-relevant meaning under practical constraints. Therefore, future benchmarks should evaluate whether the complete pipeline supports reliable task-oriented speech interaction. Metrics with task-relevant success rates are essential for assessing end-to-end pipeline reliability.
\end{itemize}

Overall, these lessons suggest that future evaluation protocols should move from isolated module testing toward integrated, system-level benchmarking. Such benchmarks should jointly report spatial accuracy, enhancement quality, recognition performance, latency, noise robustness, and semantic reliability under consistent acoustic conditions. This evaluation perspective naturally motivates a closer examination of how existing systems integrate SSL, DSE, and ASR in practice.

In the next section, we will analyze representative integrated spatial speech perception pipelines developed in recent years. Building on the evaluation criteria discussed in this section, we examine how existing methods couple spatial sensing, directional enhancement, and speech recognition through multitask learning, DOA-guided enhancement, end-to-end recognition-driven optimization, and low-latency enhancement-oriented designs.

\section{Analysis of Integrated Spatial Speech Perception Pipelines}
\label{sec6}

While sound source localization (SSL), directional speech enhancement (DSE), and automatic speech recognition (ASR) have each advanced significantly, their integration remains a key challenge for spatial speech perception systems. In realistic multi-speaker acoustic environments, a system must not only identify the spatial origin of a target speaker, but also exploit this spatial information to selectively extract and accurately recognize the corresponding speech under noise, reverberation, and interference. However, many existing systems still treat localization, enhancement, and recognition as loosely coupled modules. In such designs, localization is often used mainly for spatial awareness, enhancement is optimized according to signal-level criteria, and recognition is performed either on raw microphone signals or on enhanced signals without explicit feedback to the preceding stages. This modular separation limits the effective use of spatial cues and creates a mismatch between intermediate signal-level objectives and final system-level objectives such as recognition accuracy, semantic reliability, and interaction latency.

To address this limitation, recent work has begun to explore integrated pipelines that explicitly connect localization, speech separation or enhancement, and recognition. These approaches differ in how spatial information is represented, how DOA estimates are used, and how tightly the components are coupled during training and inference. In this section, we group representative approaches into three categories: (i) DOA and separation learning frameworks, which incorporate spatial cues into separation models through multitask or DOA-guided learning strategies; (ii) localization-aware ASR and embodied perception pipelines, which explicitly connect spatial localization, enhancement, and recognition either through unified neural optimization or through system-level integration with active sensing; and (iii) real-time enhancement-oriented pipelines, which emphasize low-latency spatial enhancement and practical deployment, but often do not include full recognition-level optimization. We review these representative approaches and analyze their design trade-offs in terms of spatial interpretability, semantic reliability, latency, and robustness.

\subsection{DOA and Separation Learning Frameworks}

A representative approach to incorporating spatial information through joint learning is MSDET \cite{MSDET}, which formulates speech separation and DOA estimation as a multitask problem. The model is typically built upon a multichannel separation network operating in the time--frequency domain, augmented with an additional DOA estimation branch, and trained using a weighted combination of separation and localization losses. In this design, the shared representation is encouraged to encode both spectral and spatial information, while separation and DOA estimation are predicted through different task-specific outputs. This provides a more integrated formulation than treating separation and localization as completely independent modules. However, the estimated DOA is not explicitly used to guide the separation process. Instead, spatial cues remain implicitly encoded within the learned latent representation, and the two outputs are predicted largely in parallel. As a result, the framework provides limited directional controllability and limited interpretability regarding how spatial estimates influence speech extraction.

To address the lack of explicit spatial control, DBNet \cite{9414187} introduces a DOA-driven beamforming architecture that more tightly integrates localization with the separation process. The model consists of an STFT front-end, a neural DOA estimation module, and parallel beamforming layers that use the estimated DOA to construct steering vectors for source-specific filtering, followed by inverse STFT to reconstruct time-domain signals. The entire system is trained end-to-end using separation objectives, eliminating the need for ground-truth DOA annotations while allowing the network to learn localization cues that benefit separation. This design improves interpretability because spatial information directly influences the beamforming operation. It also provides a clearer spatially grounded information flow: acoustic observations are first mapped to spatial estimates, which are then used to extract target speech. Nevertheless, the pipeline remains largely sequential, with localization acting as an upstream module. Separation performance is therefore sensitive to DOA estimation errors, and limited feedback is available from the enhancement stage to refine localization.

Building upon these designs, JointNet \cite{11329501} further strengthens the coupling between localization and enhancement through a joint learning framework with bidirectional interaction. The architecture consists of dedicated DOA estimation blocks and speech enhancement blocks connected through interaction modules. These modules allow spatial information to be injected into the enhancement process, while enhanced speech representations are fed back to refine DOA estimation. This mutual exchange enables the model to better exploit the synergy between spatial and spectral information, leading to improved robustness in noisy and reverberant environments. Compared with sequential or loosely coupled approaches, JointNet provides a more unified treatment of localization and enhancement. From a system-level spatial speech perception perspective, this is important because it reduces the gap between spatial sensing and signal extraction. However, the increased coupling also introduces higher computational complexity, and the framework still focuses primarily on front-end signal processing without incorporating downstream ASR objectives. Therefore, although JointNet improves the integration between SSL and DSE, it does not fully optimize the complete pathway from spatial sensing to speech recognition.

\subsection{Localization-Aware ASR and Embodied Perception Pipelines}

Directional ASR (D-ASR) \cite{9414243} represents a more complete integration strategy by jointly incorporating localization, separation, and recognition within a single end-to-end architecture. Unlike approaches where spatial information is either implicitly learned or used only to guide signal extraction, D-ASR treats the direction of arrival (DOA) of each source as an intermediate representation that directly influences the subsequent processing stages. The architecture begins with a localization subnetwork operating on multichannel phase features. A convolutional module extracts spatial representations, followed by a masking mechanism and temporal aggregation to produce source-specific features. These features are then used to estimate discrete DOA distributions for each speaker, which are converted into steering vectors and time--frequency masks. A beamforming module applies these spatial filters to separate the sources, and the resulting signals are passed to an end-to-end ASR backend, typically optimized using a CTC-based objective. Importantly, the entire system can be trained using recognition loss, reducing the need for explicit ground-truth DOA annotations or clean separation targets.

This design offers several advantages for spatial speech perception. First, by explicitly modeling source location within the network, D-ASR provides greater interpretability than fully implicit separation models. Second, the use of DOA estimates as intermediate spatial variables allows the model to build a more consistent spatial representation across frequency bands. Third, optimizing the pipeline with respect to ASR performance aligns intermediate processing with the final recognition objective. This is particularly important because signal-level improvement does not always guarantee better recognition or semantic reliability. A recognition-driven objective encourages the front-end localization and separation stages to preserve information that is useful for transcription, rather than only optimizing signal reconstruction quality.

However, D-ASR also illustrates the remaining challenges in fully integrated spatial speech perception. The model typically assumes a fixed number of speakers and relatively stable source directions over an utterance, which may limit its applicability in dynamic multi-speaker scenarios. In addition, the beamforming stage relies on classical formulations rather than fully learned spatial filtering, which may reduce robustness in highly noisy or reverberant environments. The framework also does not explicitly target real-time deployment, and the computational cost of jointly performing localization, separation, and recognition may be substantial. Therefore, D-ASR demonstrates an important step toward system-level pipeline optimization, but further work is required to improve scalability, streaming capability, and robustness to dynamic acoustic scenes.

Beyond fully neural localization-aware ASR, recent work has also explored embodied pipeline-level integration, where spatial localization is used not only to guide signal processing but also to control the physical sensing configuration of the robot. A representative example is the robotic-arm-based speech enhancement system proposed in \cite{turcotte2026lendearspeechenhancement}. In this work, a sixteen-channel microphone array is mounted across the joints of a Kinova manipulator, allowing the sensing geometry to be physically reconfigured according to the target speaker position. The pipeline first estimates an approximate target direction using an SRP-PHAT-based SSL module supported by DNN-estimated ideal ratio masks. Since the articulated arm introduces large and variable microphone spacings, as well as possible acoustic obstruction from the robot structure, this initial localization is treated as approximate rather than final. The system then uses an RGB camera and depth sensor to refine the target speaker position, after which an inverse-kinematics module moves the arm into an optimized listening configuration. Speech enhancement is subsequently performed using an MVDR beamformer whose speech and noise spatial covariance matrices are estimated from DNN-predicted masks, and the enhanced signal is evaluated using both SI-SDR and WER with a Whisper recognizer.

This embodied pipeline demonstrates a distinct form of localization-aware integration. Compared with several static microphone-array configurations and a shotgun microphone, the optimized robotic array achieves improved signal quality and lower recognition error, showing that active sensor repositioning can complement conventional spatial filtering. From the perspective of spatial speech perception, this is important because the robot body itself becomes part of the auditory front end: localization guides motion, motion improves the acoustic input, and enhancement and recognition are then performed on signals captured from a more favorable configuration. However, unlike D-ASR, this framework does not jointly optimize localization, enhancement, and recognition using an ASR loss. Its improvement should therefore be interpreted as the combined benefit of physical repositioning, improved reference-channel quality, and MVDR-based enhancement, rather than a fully recognition-driven end-to-end pipeline. This highlights an important trade-off between neural end-to-end integration and embodied system-level integration: the former aligns intermediate processing with recognition objectives, while the latter improves perception by actively changing how acoustic information is captured.

\subsection{Real-Time Enhancement-Oriented Pipelines}

In addition to integrated and end-to-end frameworks, real-time feasibility is a critical requirement for spatial speech perception systems. A representative example is a real-time stereo speech enhancement framework based on a dual-path structure with spatial-cue preservation \cite{10447153}. This system combines classical beamforming with a pretrained monaural enhancement network, such as PercepNet, to process multichannel inputs under strict latency constraints. The pipeline first performs spatial separation using beamforming driven by steering vectors, which are iteratively updated using enhanced signals to track source locations. Each separated stream is then enhanced using a shared-band gain strategy, ensuring that both channels are processed consistently to preserve interaural level and phase differences. The final output is reconstructed by remixing the enhanced spatial components, allowing the system to maintain perceptual spatial cues while improving speech quality. The overall system achieves a real-time processing latency of approximately 20--30 ms, making it suitable for interactive speech applications.

The use of lightweight neural enhancement modules and frame-based processing enables low-latency operation, showing the feasibility of practical real-time spatial enhancement. From a deployment-oriented perspective, this type of system is valuable because it directly addresses latency, which is often neglected in more complex end-to-end models. However, these benefits come with limitations. The method assumes a small and fixed number of speakers, typically two, and may be difficult to extend to more complex multi-speaker environments. It also relies on steering vector estimation rather than a fully learned spatial representation, which may reduce robustness when the acoustic scene is highly reverberant or when spatial estimates are inaccurate. Moreover, the system focuses on signal enhancement and does not incorporate downstream ASR or semantic decoding. As a result, while it demonstrates the feasibility of low-latency spatial enhancement, it also highlights the gap between practical real-time enhancement and fully integrated localization-to-recognition spatial speech perception pipelines.

\subsection{Lessons Learned}

The representative pipelines reviewed above show that integration can occur at different levels, ranging from multitask front-end learning to recognition-driven end-to-end optimization. Several lessons can be drawn from these designs.

First, simply predicting DOA and separated speech within the same network does not necessarily guarantee strong functional coupling. Multitask systems such as MSDET encourage shared spatial-spectral representations, but if DOA estimates are not explicitly used to guide separation, spatial information remains difficult to interpret and control. For spatial speech perception, integration should therefore be evaluated not only by the number of tasks included in the model, but also by how information flows between them.

Second, DOA-guided enhancement provides a more interpretable integration mechanism. Systems such as DBNet show that estimated spatial information can directly influence beamforming and source extraction. This makes the pipeline more transparent and better aligned with the system-level objective of extracting a target speaker from interference. However, these systems are also sensitive to localization errors, indicating that DOA-guided enhancement should be evaluated under controlled DOA perturbations and realistic noise conditions.

Third, bidirectional interaction between localization and enhancement can improve robustness. Joint frameworks such as JointNet demonstrate that enhanced speech can help refine DOA estimation, while DOA estimates can guide enhancement. This mutual interaction better reflects the interdependence between spatial sensing and signal extraction. Nevertheless, stronger coupling often increases model complexity, making latency and computational efficiency important concerns.

Fourth, recognition-driven optimization provides the clearest link between front-end spatial processing and downstream speech understanding. D-ASR illustrates that when the entire pipeline is optimized according to recognition loss, intermediate representations can become more aligned with the final recognition or task-oriented objective. This is especially important because signal-level metrics such as SI-SDR or PESQ may not always correlate with WER or task success. Embodied systems further show that spatial speech perception can be integrated at the level of physical sensing, where localization guides robot motion to improve the acoustic input before enhancement and recognition. However, fully unified systems often rely on simplifying assumptions, such as fixed speaker number, static DOA, or offline inference, which limits their immediate applicability to dynamic real-world scenarios.

Finally, real-time enhancement-oriented pipelines show that practical deployment constraints remain essential. Low-latency systems may not be fully integrated with ASR, but they provide valuable design principles for interactive speech systems, including causal processing, lightweight enhancement, and spatial cue preservation. Future spatial speech perception systems should combine the strengths of these directions: the interpretability of DOA-guided processing, the robustness of joint localization enhancement learning, the downstream alignment of recognition-driven optimization, and the efficiency of real-time enhancement pipelines.

\section{Applications, Challenges, and Future Directions for Spatial Speech Perception Systems}
\label{sec7}

The previous sections reviewed the foundations, evaluation criteria, and integrated architectures of spatial speech perception systems. These systems transform multichannel acoustic observations into spatial information, enhanced target speech, and linguistic or task-level outputs. The central concern in spatial speech perception is how acoustic sensing, enhancement, and recognition can be combined to support reliable speech understanding in real-world environments. This perspective is closely aligned with application domains such as hearing aids, robot audition, smart speakers, and meeting transcription.

In these applications, speech is rarely captured under clean, close-talking, and single-speaker conditions. Instead, the system must process distant speech, reverberation, background noise, overlapping speakers, moving sources, and device-dependent microphone characteristics. The practical value of SSL, DSE, and ASR therefore depends not only on component-level accuracy, but also on whether the complete pipeline can improve intelligibility, recognition reliability, target-speaker consistency, and interaction latency. This section discusses representative applications, followed by key challenges and future directions for deployable spatial speech perception systems.

\subsection{Representative Applications}

\subsubsection{Hearing Aids and Assistive Listening Devices}

Hearing aids and assistive listening devices are among the most direct application domains for spatial speech perception. Listeners with hearing impairment often experience difficulty understanding speech in noisy and multi-talker environments, which motivates the use of spatial filtering and microphone-array processing to improve the target-to-interference ratio. Classical hearing-aid front ends commonly rely on directional microphones, adaptive beamforming, multichannel Wiener filtering, and binaural noise reduction techniques to suppress interfering sources while preserving the target speech signal \cite{Doclo2010HearingAidBeamforming,van_den_bogaert2009speech}. From the perspective of spatial speech perception, these systems can be interpreted as compact low-latency pipelines in which spatial sensing and directional enhancement are used to support human speech understanding.

A distinctive requirement in hearing assistance is that enhancement should not only improve signal quality, but also preserve spatial auditory cues. In binaural hearing aids, interaural time differences, interaural level differences, and interaural coherence provide important information for sound localization and spatial awareness. Noise reduction methods that aggressively suppress interference may distort these cues, leading to an unnatural or spatially inconsistent auditory scene. This has motivated binaural enhancement methods that explicitly trade off noise reduction, speech distortion, and spatial cue preservation \cite{vandenbogaert2007binaural,marquardt2015interaural}. More recent work on cooperative and augmented listening further extends this idea by combining wearable devices with distributed microphones in the environment, suggesting that future hearing assistance may involve both local microphones and external spatial sensors \cite{corey2019cooperative,corey2020binaural}. These requirements make hearing aids an important benchmark for evaluating whether spatial speech perception systems can remain perceptually useful under strict latency, power, and form-factor constraints.

For hearing assistance, a spatial speech perception pipeline can use multichannel microphone inputs to estimate the target direction, enhance speech from that direction, and suppress spatially separated noise or competing speakers. This makes hearing aids and assistive listening devices a demanding application scenario, where real-time operation, device size, and power consumption constraints must be carefully addressed.

\subsubsection{Robot Audition and Human--Robot Interaction}

Robot audition is another central application of spatial speech perception. A robot operating in a human environment must determine where a sound is coming from, separate target speech from other sound sources, and recognize spoken instructions under ego-noise, room reverberation, and source movement. Unlike fixed smart speakers, mobile robots introduce additional challenges due to changing microphone positions, self-generated motor noise, head or body motion, and the need to associate speech with people and actions in the physical environment \cite{argentieri2015survey,Nakadai2020RobotAuditionCASA}.

Early robot audition systems already emphasized the integration of localization, separation, and recognition. For example, HARK was designed as an open-source robot audition system with modules for sound source localization, sound source separation, and automatic speech recognition of separated speech signals \cite{nakadai2010hark}. Similarly, ManyEars provided a real-time open framework for microphone-array-based localization, tracking, and separation for mobile robot audition \cite{grondin2013manyears}. More recently, ODAS has focused on embedded robot audition by reducing computational load and enabling localization, tracking, and separation on low-cost computing platforms \cite{51_grondin2022odasopenembeddedaudition}. These software frameworks are highly relevant to spatial speech perception because they demonstrate that the field is not only concerned with isolated algorithms, but also with deployable auditory pipelines that must operate in real time. Some of these existing systems have already demonstrated the capability to perform sound source localization and separation using predicted DoA information. However, more comprehensive pipelines should be further investigated, particularly in light of recent advances in deep neural networks.

For human--robot interaction, spatial speech perception enables several capabilities: identifying the active speaker, maintaining attention toward a target user, following spoken commands, detecting interruptions, and combining acoustic cues with visual perception. However, existing systems often evaluate SSL, enhancement, and ASR separately. A new research direction is therefore to evaluate robot audition as a complete interaction pipeline, reporting not only angular error or enhancement quality, but also command recognition accuracy, target-speaker consistency, response latency, and robustness under robot motion.

\subsubsection{Smart Speakers, Smart Homes, and Voice-Controlled Interfaces}

Smart speakers and voice-controlled home systems are practical examples of far-field speech perception. These devices must process speech from several meters away while handling reverberation, background television or music playback, household noise, and multiple speakers. Far-field ASR literature shows that distant speech recognition commonly requires a front end for dereverberation, source separation, and beamforming, together with a robust ASR back end trained under diverse acoustic conditions \cite{HaebUmbach2021FarFieldASR}. Microphone arrays are therefore a key component of smart-speaker front ends, supporting source localization, beamforming, acoustic echo cancellation, and noise suppression before wake-word detection or full ASR \cite{chhetri2018multichannel}.

Smart-home speech interfaces further connect acoustic perception to device control. Voice-controlled smart-home frameworks typically combine ASR, natural language processing, and IoT control logic to transform spoken utterances into executable actions \cite{iliev2023framework}. In such systems, spatial speech perception is useful because the intended command may be mixed with other household sounds or conversations. SSL can estimate the active speaker location, DSE can enhance the speech from that direction, and ASR can convert the enhanced signal into a command. Domestic audio datasets such as CHiME-Home and distant multi-microphone challenges such as CHiME-5 further illustrate the acoustic complexity of home environments, including background activities, multiple speakers, and natural conversational speech \cite{foster2015chimehome,Barker2018CHiME5}. This application area is important because it directly connects spatial processing with robust ASR, user-facing interaction, and real-world acoustic deployment.

\subsubsection{Teleconferencing and Meeting Transcription}

Teleconferencing and meeting transcription are natural applications for spatial speech perception because they involve distant microphones, overlapping speech, room reverberation, and multiple participants. Meeting corpora such as AMI provide multimodal meeting recordings with transcriptions and annotations, supporting research on speech recognition, diarization, and meeting understanding \cite{carletta2005ami}. More recent datasets and challenges, including CHiME-5 and LibriCSS, explicitly target distant multi-microphone conversational ASR and continuous speech separation in meeting-like scenarios \cite{Barker2018CHiME5,chen2020libricss}. These datasets are especially relevant to integrated spatial speech perception because they require the system to handle both overlap and continuity rather than isolated pre-segmented mixtures.

In meeting transcription, SSL can support speaker localization and diarization, DSE can suppress interfering speakers or room noise, and ASR can produce transcriptions for downstream summarization or indexing. However, the relationship between enhancement and recognition is not always straightforward. An enhancement model may improve signal-level metrics while introducing artifacts that degrade word recognition. Therefore, meeting transcription provides a useful test case for recognition-oriented spatial enhancement: the front end should be evaluated according to downstream WER, diarization consistency, overlap handling, latency, and robustness across microphone configurations.

\subsubsection{Wearable, Mobile, and Accessibility-Oriented Interfaces}

Wearable and mobile devices, including earbuds, smart glasses, augmented-reality headsets, and smartphones, introduce another application domain for spatial speech perception. These platforms combine strict constraints on power, latency, and memory with strong requirements for natural interaction. In wearable listening devices, spatial enhancement can improve speech intelligibility while preserving the user's awareness of the surrounding acoustic scene \cite{corey2019cooperative,corey2020binaural}. In mobile voice interfaces, wake-word detection and on-device recognition must operate continuously with low power consumption, while more complex ASR or semantic interpretation may be activated only when needed \cite{heySiri2017}.

Spatial speech perception is also relevant for accessibility-oriented systems such as real-time captioning, assistive communication, and target-speaker enhancement. In audio-visual settings, visual information can help identify the target speaker and guide speech separation when multiple people are speaking simultaneously \cite{ephrat2018looking}. This suggests that future wearable systems may combine microphone-array signals, visual cues, user gaze, and contextual information to determine which speaker should be enhanced or recognized.

\subsection{Key Challenges}

\subsubsection{Mismatch Between Signal-Level and Task-Level Objectives}

A persistent challenge is the mismatch between signal-level enhancement metrics and downstream task performance. Enhancement models are often optimized using signal reconstruction objectives or evaluated using metrics such as SI-SDR, PESQ, and STOI. However, these metrics do not always predict ASR accuracy, command success, or perceived listening benefit. In spatial speech perception, the final objective is not always to reconstruct a clean waveform, but to preserve the information required for recognition, interaction, or human listening. Future systems should therefore evaluate enhancement jointly with downstream ASR, speech intelligibility, target-speaker consistency, and application-specific task success.

\subsubsection{Robustness Under Dynamic Multi-Speaker Conditions}

Many existing SSL, DSE, and ASR methods assume a fixed number of speakers, static source positions, or short utterance-level inputs. Real-world applications violate these assumptions. Speakers may move, overlap intermittently, enter or leave the scene, or speak across chunk boundaries. In robotics and smart-home environments, the target speaker may also change depending on interaction context. This creates a need for spatial speech perception systems that can handle dynamic DOA trajectories, source counting uncertainty, intermittent speech activity, and target-speaker switching.

\subsubsection{Error Propagation Across the Pipeline}

Integrated SSL--DSE--ASR systems are sensitive to error propagation. An inaccurate DOA estimate can cause DSE to enhance the wrong direction, and enhancement artifacts can increase ASR errors. Conversely, ASR confidence or linguistic context may provide useful feedback for detecting front-end failures. Current pipelines are often feed-forward, with limited interaction between localization, enhancement, and recognition. Future research should consider uncertainty-aware spatial representations, multiple-hypothesis enhancement, confidence-based reprocessing, and closed-loop feedback from ASR to the spatial front end.

\subsubsection{Real-Time and Embedded Deployment Constraints}

Many applications require low and predictable latency. Hearing aids, robot audition, smart speakers, and wearable interfaces cannot rely on long offline processing windows or computationally expensive models. Real-time systems must balance spatial accuracy, enhancement quality, recognition performance, memory usage, power consumption, and hardware constraints. This motivates causal architectures, lightweight neural models, streaming ASR, efficient beamforming, and hybrid approaches that combine interpretable signal processing with learning-based robustness.

\subsubsection{Benchmarking Across Complete Pipelines}

A major limitation in the literature is the lack of standardized benchmarks for complete spatial speech perception systems. SSL, DSE, and ASR are often evaluated using different datasets and assumptions, making it difficult to determine whether improvements in one component improve the full system. Future benchmarks should include multichannel audio, realistic reverberation and noise, multiple and moving speakers, DOA annotations, clean references, transcriptions, and task-level labels. Evaluation should report angular error, enhancement quality, intelligibility, WER, diarization performance, latency, computational cost, and robustness under controlled acoustic perturbations.

\subsection{Future Directions}

Future spatial speech perception systems should move from component-level optimization toward perception-aware and task-oriented design. First, recognition-oriented enhancement should be further explored so that DSE is optimized not only for waveform reconstruction but also for downstream ASR and command understanding. Directional ASR and jointly optimized localization--separation--recognition models provide promising examples of this direction \cite{9414243,HaebUmbach2021FarFieldASR}.

Second, future systems should incorporate uncertainty-aware spatial processing. Instead of passing a single DOA estimate from SSL to DSE, the system may pass a probability distribution over directions, multiple candidate sources, or confidence-aware spatial embeddings. This would allow the enhancement and recognition stages to remain robust when localization is ambiguous or unstable.

Third, real-time deployment should become a first-class evaluation criterion. Systems intended for hearing aids, smart speakers, robots, and wearable devices should report end-to-end latency, real-time factor, memory usage, and hardware platform, in addition to accuracy metrics. This is particularly important for interactive applications where delayed responses can reduce usability even when recognition accuracy is high.

Fourth, multimodal scene-aware spatial speech perception is likely to become increasingly important. In robotics, augmented reality, and wearable interfaces, audio can be combined with vision, gaze, face tracking, gesture recognition, and scene understanding to identify the intended speaker and interaction context. Vision-based perception has demonstrated strong capability in human tracking \cite{Milan2016MOT,Chen2017DeepTracking}, gesture recognition \cite{Molchanov2016Gesture}, and visual scene understanding \cite{Redmon2016YOLO}. In addition, recent audio--visual perception frameworks have shown that combining spatial hearing with visual information can improve scene understanding by exploiting cross-modal consistency between visual motion cues and acoustic spatial cues \cite{gan2020looklistenactaudiovisual,chen2020soundspacesaudiovisualnavigation3d,chen2021learningsetwaypointsaudiovisual}. Future systems should integrate multimodal cues with SSL, DSE, and ASR in a temporally consistent and deployable pipeline.

Finally, software reproducibility and real-world testing should be strengthened. Open auditory middleware, simulation tools, and benchmark datasets should support the same pipeline across offline simulation, recorded real-world data, and live deployment. This would make it easier to compare SSL--DSE--ASR systems under realistic constraints and accelerate progress toward robust, low-latency, and application-aware spatial speech perception.

\section{Conclusion}
\label{sec8}

This paper has presented a survey of spatial speech perception systems, with particular emphasis on the integration of sound source localization (SSL), directional speech enhancement (DSE), and automatic speech recognition (ASR). Motivated by the growing need for robust speech understanding in realistic acoustic environments, we considered spatial speech perception as a complete processing framework that transforms multichannel acoustic observations into spatial estimates, enhanced target speech, and linguistic outputs. This perspective is relevant to a wide range of speech and audio applications, including hearing aids, assistive listening devices, robot audition, smart speakers, teleconferencing, meeting transcription, wearable interfaces, and intelligent voice-controlled systems.

We first reviewed the foundations of SSL, DSE, and ASR. For SSL, we discussed traditional signal-processing methods, including TDoA, GCC-PHAT, SRP-PHAT, MVDR-based localization, and subspace-based approaches, together with learning-based methods based on CNNs, CRNNs, and attention mechanisms. For DSE, we summarized filtering-based, masking-based, and DOA-conditioned approaches, highlighting how spatial information can be used to enhance target speech and suppress competing sources. For ASR, we reviewed the development from traditional hybrid GMM-HMM and DNN-HMM systems to modern end-to-end, Transformer-based, Conformer-based, self-supervised, and large-scale pretrained models. These component-level developments show substantial progress, but they also reveal that robust real-world speech perception cannot be achieved by optimizing localization, enhancement, or recognition in isolation.

A central message of this survey is that spatial speech perception should be evaluated at the system level. In practical auditory systems, errors and delays propagate across the SSL--DSE--ASR pipeline. An inaccurate or unstable DOA estimate may cause the enhancement module to suppress the target speaker or enhance an interfering source. Similarly, an enhancement model that improves signal-level quality may introduce artifacts that degrade recognition performance, while an ASR model with strong offline accuracy may still be unsuitable for interactive applications if it requires long input contexts or produces delayed outputs. Therefore, evaluation should jointly consider spatial accuracy, enhancement quality, intelligibility, recognition performance, target-speaker consistency, latency, computational efficiency, and robustness to noise, reverberation, and overlapping speech.

We further reviewed representative integrated pipelines that connect localization, enhancement, separation, and recognition. Existing systems demonstrate a clear trend from loosely coupled modular designs toward more integrated architectures, including DOA-guided enhancement, multitask localization and separation, bidirectional localization--enhancement interaction, and recognition-oriented optimization. Nevertheless, several challenges remain unresolved. Current systems are often evaluated under simplified assumptions about the number of speakers, source movement, array configuration, and acoustic conditions. Real-time performance is also not consistently reported, and standardized benchmarks for complete SSL--DSE--ASR pipelines remain limited. These limitations make it difficult to assess whether improvements in one component reliably translate into better end-to-end speech understanding.

The application analysis further shows that different deployment scenarios impose different requirements on spatial speech perception systems. Hearing aids and assistive listening devices require low-latency, noise-robust, and perceptually meaningful enhancement under strict power and form-factor constraints. Robot audition requires spatially aware speech perception under ego-noise, movement, and dynamic human--robot interaction. Smart speakers and smart-home interfaces require robust far-field speech capture in reverberant domestic environments, while meeting transcription and teleconferencing systems must handle overlapping speakers and long-form conversational speech. These applications highlight the need for spatial speech perception systems that are not only accurate, but also robust, efficient, adaptive, and suitable for real-world deployment.

Future research should therefore move toward perception-aware and task-oriented spatial speech processing. Promising directions include recognition-oriented enhancement, uncertainty-aware DOA estimation, joint localization--enhancement--recognition modeling, streaming and causal architectures, multimodal scene-aware perception, and standardized pipeline-level benchmarks. Greater attention should also be given to real-time implementation, resource-constrained deployment, multichannel microphone requirements, and evaluation under realistic acoustic conditions. By jointly considering spatial sensing, target enhancement, recognition reliability, latency, and robustness, future spatial speech perception systems can better support reliable speech understanding in complex real-world environments.

\clearpage

\bibliographystyle{IEEEtran}
\bibliography{references}

@article{Nakadai2020RobotAuditionCASA,
    author = {Nakadai, Kazuhiro and Okuno, Hiroshi G.},
    title = {Robot Audition and Computational Auditory Scene Analysis},
    journal = {Advanced Intelligent Systems},
    volume = {2},
    number = {9},
    pages = {2000050},
    doi = {https://doi.org/10.1002/aisy.202000050},
    url = {https://advanced.onlinelibrary.wiley.com/doi/abs/10.1002/aisy.202000050},
    eprint = {https://advanced.onlinelibrary.wiley.com/doi/pdf/10.1002/aisy.202000050},
    year = {2020}
}

@article{Jekaterynczuk2024SSLSurvey,
  title={A Survey of Sound Source Localization and Detection Methods and Their Applications},
  author={Jekatery{\'n}czuk, G. and others},
  journal={Sensors},
  year={2023},
  url={https://pmc.ncbi.nlm.nih.gov/articles/PMC10781166/}
}

@book{BrandsteinSilvermanBook,
  title={Microphone Arrays: Signal Processing Techniques and Applications},
  author={Brandstein, Michael and Silverman, Harvey},
  publisher={Springer},
  year={2001}
}

@article{Grumiaux2022SurveyDeepSSL,
  title={A Survey of Sound Source Localization with Deep Learning Methods},
  author={Grumiaux, Pierre-Amaury and others},
  journal={The Journal of the Acoustical Society of America},
  year={2022},
  volume={152},
  number={1},
  pages={107--136},
  url={https://pubs.aip.org/asa/jasa/article/152/1/107/2838290/A-survey-of-sound-source-localization-with-deep}
}

@inproceedings{Baevski2020Wav2vec2,
  title={wav2vec 2.0: A Framework for Self-Supervised Learning of Speech Representations},
  author={Baevski, Alexei and others},
  booktitle={NeurIPS},
  year={2020},
  url={https://arxiv.org/abs/2006.11477}
}

@article{Knapp1976GCC,
  title        = {The Generalized Correlation Method for Estimation of Time Delay},
  author       = {Knapp, Charles and Carter, G. Clifford},
  journal      = {IEEE Transactions on Acoustics, Speech, and Signal Processing},
  year         = {1976},
  volume       = {24},
  number       = {4},
  pages        = {320--327},
  doi          = {10.1109/TASSP.1976.1162830}
}

@inproceedings{Rui2004TDE,
  title        = {Time Delay Estimation in the Presence of Correlated Noise and Reverberation},
  author       = {Rui, Yong and Florencio, Dinei},
  booktitle    = {Proceedings of IEEE International Conference on Acoustics, Speech, and Signal Processing (ICASSP)},
  year         = {2004},
  doi          = {10.1109/ICASSP.2004.1326212}

}

@inproceedings{Silverman1997RobustTDE,
  title        = {A Robust Method for Speech Signal Time-Delay Estimation in the Presence of Reverberation and Noise},
  author       = {Silverman, H. F. and Yu, Y. and Sachar, J. and Patterson, W.},
  booktitle    = {Proceedings of IEEE International Conference on Acoustics, Speech, and Signal Processing (ICASSP)},
  year         = {1997}
}

@article{TDoA,
author = {Xu, Bin and Sun, Guodong and Yu, Ran and Yang, Zheng},
year = {2013},
month = {08},
pages = {1567-1576},
title = {High-Accuracy TDOA-Based Localization without Time Synchronization},
volume = {24},
journal = {Parallel and Distributed Systems, IEEE Transactions on},
doi = {10.1109/TPDS.2012.248}
}

@book{delay-and-sum,
  author    = {Van Trees, Harry L.},
  title     = {Optimum Array Processing: Part IV of Detection, Estimation, and Modulation Theory},
  year      = {2002},
  publisher = {Wiley},
  address   = {New York},
  isbn      = {9780471093909}
}

@ARTICLE{MVDR_1449208,
  author={Capon, J.},
  journal={Proceedings of the IEEE}, 
  title={High-resolution frequency-wavenumber spectrum analysis}, 
  year={1969},
  volume={57},
  number={8},
  pages={1408-1418},
  doi={10.1109/PROC.1969.7278}}

@article{MUSIC_schmidt1986music,
  author    = {Schmidt, R. O.},
  title     = {Multiple Emitter Location and Signal Parameter Estimation},
  journal   = {IEEE Transactions on Antennas and Propagation},
  volume    = {34},
  number    = {3},
  pages     = {276--280},
  year      = {1986}
}

@article{WangKaveh1985Broadband,
  title={Coherent signal-subspace processing for the detection and estimation of angles of arrival of multiple wide-band sources},
  author={Wang, H. and Kaveh, M.},
  journal={IEEE Transactions on Acoustics, Speech, and Signal Processing},
  year={1985}
}

@INPROCEEDINGS{GEVD_MUSIC_5354419,
  author={Nakamura, Keisuke and Nakadai, Kazuhiro and Asano, Futoshi and Hasegawa, Yuji and Tsujino, Hiroshi},
  booktitle={2009 IEEE/RSJ International Conference on Intelligent Robots and Systems}, 
  title={Intelligent sound source localization for dynamic environments}, 
  year={2009},
  volume={},
  number={},
  pages={664-669},
  doi={10.1109/IROS.2009.5354419}}

@INPROCEEDINGS{GSVD_6385494,
  author={Nakamura, Keisuke and Nakadai, Kazuhiro and Ince, Gökhan},
  booktitle={2012 IEEE/RSJ International Conference on Intelligent Robots and Systems}, 
  title={Real-time super-resolution Sound Source Localization for robots}, 
  year={2012},
  volume={},
  number={},
  pages={694-699},
  doi={10.1109/IROS.2012.6385494}}

@INPROCEEDINGS{IGSVD_6942813,
  author={Ohata, Takuma and Nakamura, Keisuke and Mizumoto, Takeshi and Taiki, Tezuka and Nakadai, Kazuhiro},
  booktitle={2014 IEEE/RSJ International Conference on Intelligent Robots and Systems}, 
  title={Improvement in outdoor sound source detection using a quadrotor-embedded microphone array}, 
  year={2014},
  volume={},
  number={},
  pages={1902-1907},
  doi={10.1109/IROS.2014.6942813}}

@inproceedings{Chakrabarty_2017,
   title={Broadband doa estimation using convolutional neural networks trained with noise signals},
   url={http://dx.doi.org/10.1109/WASPAA.2017.8170010},
   DOI={10.1109/waspaa.2017.8170010},
   booktitle={2017 IEEE Workshop on Applications of Signal Processing to Audio and Acoustics (WASPAA)},
   publisher={IEEE},
   author={Chakrabarty, Soumitro and Habets, Emanuel A. P.},
   year={2017},
   month=oct, pages={136–140} }

@article{Milan2016MOT,
  author  = {Milan, Anton and Leal-Taix{\'e}, Laura and Reid, Ian and Roth, Stefan and Schindler, Konrad},
  title   = {MOT16: A benchmark for multi-object tracking},
  journal = {arXiv preprint arXiv:1603.00831},
  year    = {2016}
}

@article{Chen2017DeepTracking,
  author  = {Chen, Long and Ai, Haizhou and Zhuang, Zijie and Shang, Chong},
  title   = {Real-time multiple people tracking with deeply learned candidate selection and person re-identification},
  journal = {IEEE International Conference on Multimedia and Expo (ICME)},
  year    = {2017}
}

@inproceedings{Molchanov2016Gesture,
  author    = {Molchanov, Pavlo and Yang, Xiaodong and Gupta, Shalini and Kim, Kihwan and Tyree, Stephen and Kautz, Jan},
  title     = {Online detection and classification of dynamic hand gestures with recurrent 3D convolutional neural networks},
  booktitle = {IEEE Conference on Computer Vision and Pattern Recognition (CVPR)},
  year      = {2016}
}

@inproceedings{Redmon2016YOLO,
  author    = {Redmon, Joseph and Divvala, Santosh and Girshick, Ross and Farhadi, Ali},
  title     = {You Only Look Once: Unified, real-time object detection},
  booktitle = {IEEE Conference on Computer Vision and Pattern Recognition (CVPR)},
  year      = {2016}
}

@misc{gan2020looklistenactaudiovisual,
      title={Look, Listen, and Act: Towards Audio-Visual Embodied Navigation}, 
      author={Chuang Gan and Yiwei Zhang and Jiajun Wu and Boqing Gong and Joshua B. Tenenbaum},
      year={2020},
      eprint={1912.11684},
      archivePrefix={arXiv},
      primaryClass={cs.CV},
      url={https://arxiv.org/abs/1912.11684}, 
}

@misc{chen2020soundspacesaudiovisualnavigation3d,
      title={SoundSpaces: Audio-Visual Navigation in 3D Environments}, 
      author={Changan Chen and Unnat Jain and Carl Schissler and Sebastia Vicenc Amengual Gari and Ziad Al-Halah and Vamsi Krishna Ithapu and Philip Robinson and Kristen Grauman},
      year={2020},
      eprint={1912.11474},
      archivePrefix={arXiv},
      primaryClass={cs.CV},
      url={https://arxiv.org/abs/1912.11474}, 
}

@misc{chen2021learningsetwaypointsaudiovisual,
      title={Learning to Set Waypoints for Audio-Visual Navigation}, 
      author={Changan Chen and Sagnik Majumder and Ziad Al-Halah and Ruohan Gao and Santhosh Kumar Ramakrishnan and Kristen Grauman},
      year={2021},
      eprint={2008.09622},
      archivePrefix={arXiv},
      primaryClass={cs.CV},
      url={https://arxiv.org/abs/2008.09622}, 
}

@inproceedings{1_He_2018,
   title={Deep Neural Networks for Multiple Speaker Detection and Localization},
   url={http://dx.doi.org/10.1109/ICRA.2018.8461267},
   DOI={10.1109/icra.2018.8461267},
   booktitle={2018 IEEE International Conference on Robotics and Automation (ICRA)},
   publisher={IEEE},
   author={He, Weipeng and Motlicek, Petr and Odobez, Jean-Marc},
   year={2018},
   month=may, pages={74–79} }

@article{2_akter2025hybrid,
  title        = {A hybrid {CNN}-{LSTM} model for environmental sound classification: Leveraging feature engineering and transfer learning},
  author       = {Akter, Rubaiya and Islam, Md. Rezwanul and Debnath, Sumon Kumar and Sarker, Prodip Kumar and Uddin, Md. Kamal},
  journal      = {Digital Signal Processing},
  volume       = {163},
  pages        = {105234},
  year         = {2025},
  issn         = {1051-2004},
  doi          = {10.1016/j.dsp.2025.105234},
  url          = {https://www.sciencedirect.com/science/article/pii/S1051200425002568}
}

@article{3_zhang2025multiple,
  title        = {Multiple Sound Sources Localization Using Sub-Band Spatial Features and Attention Mechanism},
  author       = {Zhang, D. and Chen, J. and Bai, J. and others},
  journal      = {Circuits, Systems, and Signal Processing},
  volume       = {44},
  pages        = {2592--2620},
  year         = {2025},
  doi          = {10.1007/s00034-024-02925-6},
  url          = {https://doi.org/10.1007/s00034-024-02925-6},
  note         = {Published 13 December 2024, Issue date April 2025}
}

@article{4_Shin2023SoundEL,
  title={Sound Event Localization and Detection Using Imbalanced Real and Synthetic Data via Multi-Generator},
  author={Yeong Cheol Shin and Chanjun Chun},
  journal={Sensors (Basel, Switzerland)},
  year={2023},
  volume={23},
  url={https://api.semanticscholar.org/CorpusID:257745516}
}

@article{5_boztas2023sound,
  title        = {Sound source localization for auditory perception of a humanoid robot using deep neural networks},
  author       = {Boztas, G.},
  journal      = {Neural Computing and Applications},
  volume       = {35},
  pages        = {6801--6811},
  year         = {2023},
  doi          = {10.1007/s00521-022-08047-x},
  url          = {https://doi.org/10.1007/s00521-022-08047-x},
  note         = {Published 29 November 2022, Issue date March 2023}
}

@misc{6_liu2025soundsourcelocalizationhumanrobot,
      title={Sound Source Localization for Human-Robot Interaction in Outdoor Environments}, 
      author={Victor Liu and Timothy Du and Jordy Sehn and Jack Collier and François Grondin},
      year={2025},
      eprint={2507.21431},
      archivePrefix={arXiv},
      primaryClass={cs.RO},
      url={https://arxiv.org/abs/2507.21431}, 
}

@INPROCEEDINGS{7_8206494,
  author={Nakadai, Kazuhiro and Kumon, Makoto and Okuno, Hiroshi G. and Hoshiba, Kotaro and Wakabayashi, Mizuho and Washizaki, Kai and Ishiki, Takahiro and Gabriel, Daniel and Bando, Yoshiaki and Morito, Takayuki and Kojima, Ryosuke and Sugiyama, Osamu},
  booktitle={2017 IEEE/RSJ International Conference on Intelligent Robots and Systems (IROS)}, 
  title={Development of microphone-array-embedded UAV for search and rescue task}, 
  year={2017},
  volume={},
  number={},
  pages={5985-5990},
  doi={10.1109/IROS.2017.8206494}}

@misc{8_lagacé2023egonoisereductionmobilerobot,
      title={Ego-noise reduction of a mobile robot using noise spatial covariance matrix learning and minimum variance distortionless response}, 
      author={Pierre-Olivier Lagacé and François Ferland and François Grondin},
      year={2023},
      eprint={2303.00829},
      archivePrefix={arXiv},
      primaryClass={eess.AS},
      url={https://arxiv.org/abs/2303.00829}, 
}

@misc{9_wang2025singlemicrophonebasedsoundsourcelocalization,
      title={Single-Microphone-Based Sound Source Localization for Mobile Robots in Reverberant Environments}, 
      author={Jiang Wang and Runwu Shi and Benjamin Yen and He Kong and Kazuhiro Nakadai},
      year={2025},
      eprint={2506.16173},
      archivePrefix={arXiv},
      primaryClass={cs.RO},
      url={https://arxiv.org/abs/2506.16173}, 
}

@misc{10_wang2025ipdnet2efficientimprovedinterchannel,
      title={IPDnet2: an efficient and improved inter-channel phase difference estimation network for sound source localization}, 
      author={Yabo Wang and Bing Yang and Xiaofei Li},
      year={2025},
      eprint={2509.21900},
      archivePrefix={arXiv},
      primaryClass={eess.AS},
      url={https://arxiv.org/abs/2509.21900}, 
}

@ARTICLE{31_10319788,
  author={An, Inkyu and An, Guoyuan and Kim, Taeyoung and Yoon, Sung-eui},
  journal={IEEE Robotics and Automation Letters}, 
  title={Microphone Pair Training for Robust Sound Source Localization With Diverse Array Configurations}, 
  year={2024},
  volume={9},
  number={1},
  pages={319-326},
  doi={10.1109/LRA.2023.3333700}}

@inproceedings{32_jiang2024duet-srp-phat,
  title        = {Development of a High-Precision Multi-Source Localization System Based on DUET-SRP-PHAT},
  author       = {Jiang, Yujie and Hang, Rongzhi and Liu, Ben and Yang, Songyu and Xu, Yishen},
  booktitle    = {2024 20th International Conference on Natural Computation, Fuzzy Systems and Knowledge Discovery (ICNC-FSKD)},
  year         = {2024},
  doi          = {10.1109/icnc-fskd64080.2024.10702295},
  url          = {https://doi.org/10.1109/icnc-fskd64080.2024.10702295}
}

@INPROCEEDINGS{33_9561885,
  author={Wang, Jiadong and Qian, Xinyuan and Pan, Zihan and Zhang, Malu and Li, Haizhou},
  booktitle={2021 IEEE International Conference on Robotics and Automation (ICRA)}, 
  title={GCC-PHAT with Speech-oriented Attention for Robotic Sound Source Localization}, 
  year={2021},
  volume={},
  number={},
  pages={5876-5883},
  doi={10.1109/ICRA48506.2021.9561885}}

@misc{51_grondin2022odasopenembeddedaudition,
      title={ODAS: Open embeddeD Audition System}, 
      author={François Grondin and Dominic Létourneau and Cédric Godin and Jean-Samuel Lauzon and Jonathan Vincent and Simon Michaud and Samuel Faucher and François Michaud},
      year={2022},
      eprint={2103.03954},
      archivePrefix={arXiv},
      primaryClass={eess.AS},
      url={https://arxiv.org/abs/2103.03954}, 
}

@article{40_SELFnet,
  author       = {Sharath Adavanne and
                  Archontis Politis and
                  Joonas Nikunen and
                  Tuomas Virtanen},
  title        = {Sound Event Localization and Detection of Overlapping Sources Using
                  Convolutional Recurrent Neural Networks},
  journal      = {CoRR},
  volume       = {abs/1807.00129},
  year         = {2018},
  url          = {http://arxiv.org/abs/1807.00129},
  eprinttype    = {arXiv},
  eprint       = {1807.00129},
  bibsource    = {dblp computer science bibliography, https://dblp.org}
}

@techreport{41_Du_NERCSLIP_task3_report,
  author       = {Wang, Qing and Chai, Li and Wu, Huaxin and Nian, Zhaoxu and Niu, Shutong and Zheng, Siyuan and Wang, Yuyang and Sun, Lei and Fang, Yi and Pan, Jia and Du, Jun and Lee, Chin-Hui},
  title        = {The NERC-SLIP System for Sound Event Localization and Detection of DCASE2022 Challenge},
  institution  = {DCASE2022 Challenge},
  year         = {2022},
  month        = {June},
  note         = {Technical Report},
  url          = {https://dcase.community/documents/challenge2022/technical_reports/DCASE2022_Du_122_t3.pdf}
}

@misc{70_jazaeri2025multispeakerdoaestimationbinaural,
      title={Multi-Speaker DOA Estimation in Binaural Hearing Aids using Deep Learning and Speaker Count Fusion}, 
      author={Farnaz Jazaeri and Homayoun Kamkar-Parsi and François Grondin and Martin Bouchard},
      year={2025},
      eprint={2509.21382},
      archivePrefix={arXiv},
      primaryClass={eess.AS},
      url={https://arxiv.org/abs/2509.21382}, 
}

@INPROCEEDINGS{80_8967690,
  author={Grondin, François and Glass, James},
  booktitle={2019 IEEE/RSJ International Conference on Intelligent Robots and Systems (IROS)}, 
  title={Fast and Robust 3-D Sound Source Localization with DSVD-PHAT}, 
  year={2019},
  volume={},
  number={},
  pages={5352-5357},
  doi={10.1109/IROS40897.2019.8967690}}

@misc{81_fu2025auralnethierarchicalattentionbased3d,
      title={AuralNet: Hierarchical Attention-based 3D Binaural Localization of Overlapping Speakers}, 
      author={Linya Fu and Yu Liu and Zhijie Liu and Zedong Yang and Zhong-Qiu Wang and Youfu Li and He Kong},
      year={2025},
      eprint={2506.02773},
      archivePrefix={arXiv},
      primaryClass={eess.AS},
      url={https://arxiv.org/abs/2506.02773}, 
}

@misc{CDUNet,
      title={Neural Directed Speech Enhancement with Dual Microphone Array in High Noise Scenario}, 
      author={Wen Wen and Qiang Zhou and Yu Xi and Haoyu Li and Ziqi Gong and Kai Yu},
      year={2024},
      eprint={2412.18141},
      archivePrefix={arXiv},
      primaryClass={eess.AS},
      url={https://arxiv.org/abs/2412.18141}, 
}

@misc{Deepclustering,
      title={Deep clustering: Discriminative embeddings for segmentation and separation}, 
      author={John R. Hershey and Zhuo Chen and Jonathan Le Roux and Shinji Watanabe},
      year={2015},
      eprint={1508.04306},
      archivePrefix={arXiv},
      primaryClass={cs.NE},
      url={https://arxiv.org/abs/1508.04306}, 
}

@inproceedings{DANet,
   title={Deep attractor network for single-microphone speaker separation},
   url={http://dx.doi.org/10.1109/ICASSP.2017.7952155},
   DOI={10.1109/icassp.2017.7952155},
   booktitle={2017 IEEE International Conference on Acoustics, Speech and Signal Processing (ICASSP)},
   publisher={IEEE},
   author={Chen, Zhuo and Luo, Yi and Mesgarani, Nima},
   year={2017},
   month=mar, pages={246–250} }

@misc{DRN,
      title={All Neural Low-latency Directional Speech Extraction}, 
      author={Ashutosh Pandey and Sanha Lee and Juan Azcarreta and Daniel Wong and Buye Xu},
      year={2024},
      eprint={2407.04879},
      archivePrefix={arXiv},
      primaryClass={cs.SD},
      url={https://arxiv.org/abs/2407.04879}, 
}

@misc{FASNET,
      title={FaSNet: Low-latency Adaptive Beamforming for Multi-microphone Audio Processing}, 
      author={Yi Luo and Enea Ceolini and Cong Han and Shih-Chii Liu and Nima Mesgarani},
      year={2019},
      eprint={1909.13387},
      archivePrefix={arXiv},
      primaryClass={eess.AS},
      url={https://arxiv.org/abs/1909.13387}, 
}

@article{JNF,
   title={Insights Into Deep Non-Linear Filters for Improved Multi-Channel Speech Enhancement},
   volume={31},
   ISSN={2329-9304},
   url={http://dx.doi.org/10.1109/TASLP.2022.3221046},
   DOI={10.1109/taslp.2022.3221046},
   journal={IEEE/ACM Transactions on Audio, Speech, and Language Processing},
   publisher={Institute of Electrical and Electronics Engineers (IEEE)},
   author={Tesch, Kristina and Gerkmann, Timo},
   year={2023},
   pages={563–575} }

@misc{MIDEANet,
      title={End-to-End DOA-Guided Speech Extraction in Noisy Multi-Talker Scenarios}, 
      author={Kangqi Jing and Wenbin Zhang and Yu Gao},
      year={2025},
      eprint={2507.20926},
      archivePrefix={arXiv},
      primaryClass={eess.AS},
      url={https://arxiv.org/abs/2507.20926}, 
}

@ARTICLE{Generalized_sidelobe_canceller,
  author={Griffiths, L. and Jim, C.},
  journal={IEEE Transactions on Antennas and Propagation}, 
  title={An alternative approach to linearly constrained adaptive beamforming}, 
  year={1982},
  volume={30},
  number={1},
  pages={27-34},
  doi={10.1109/TAP.1982.1142739}}

@ARTICLE{rabiner1989tutorial,
  author={Rabiner, L.R.},
  journal={Proceedings of the IEEE}, 
  title={A tutorial on hidden Markov models and selected applications in speech recognition}, 
  year={1989},
  volume={77},
  number={2},
  pages={257-286},
  doi={10.1109/5.18626}}

@ARTICLE{hinton2012deep,
  author={Hinton, Geoffrey and Deng, Li and Yu, Dong and Dahl, George E. and Mohamed, Abdel-rahman and Jaitly, Navdeep and Senior, Andrew and Vanhoucke, Vincent and Nguyen, Patrick and Sainath, Tara N. and Kingsbury, Brian},
  journal={IEEE Signal Processing Magazine}, 
  title={Deep Neural Networks for Acoustic Modeling in Speech Recognition: The Shared Views of Four Research Groups}, 
  year={2012},
  volume={29},
  number={6},
  pages={82-97},
  doi={10.1109/MSP.2012.2205597}}

@inproceedings{sak2014long,
  title={Long short-term memory recurrent neural network architectures for large scale acoustic modeling.},
  author={Sak, Hasim and Senior, Andrew W and Beaufays, Fran{\c{c}}oise and others},
  booktitle={Interspeech},
  volume={2014},
  pages={338--342},
  year={2014}
}

@inproceedings{graves2006connectionist,
  title={Connectionist temporal classification: labelling unsegmented sequence data with recurrent neural networks},
  author={Graves, Alex and Fern{\'a}ndez, Santiago and Gomez, Faustino and Schmidhuber, J{\"u}rgen},
  booktitle={Proceedings of the 23rd international conference on Machine learning},
  pages={369--376},
  year={2006}
}

@misc{hannun2014deepspeechscalingendtoend,
      title={Deep Speech: Scaling up end-to-end speech recognition}, 
      author={Awni Hannun and Carl Case and Jared Casper and Bryan Catanzaro and Greg Diamos and Erich Elsen and Ryan Prenger and Sanjeev Satheesh and Shubho Sengupta and Adam Coates and Andrew Y. Ng},
      year={2014},
      eprint={1412.5567},
      archivePrefix={arXiv},
      primaryClass={cs.CL},
      url={https://arxiv.org/abs/1412.5567}, 
}

@misc{graves2012sequencetransductionrecurrentneural,
      title={Sequence Transduction with Recurrent Neural Networks}, 
      author={Alex Graves},
      year={2012},
      eprint={1211.3711},
      archivePrefix={arXiv},
      primaryClass={cs.NE},
      url={https://arxiv.org/abs/1211.3711}, 
}

@INPROCEEDINGS{8268935,
  author={Rao, Kanishka and Sak, Haşim and Prabhavalkar, Rohit},
  booktitle={2017 IEEE Automatic Speech Recognition and Understanding Workshop (ASRU)}, 
  title={Exploring architectures, data and units for streaming end-to-end speech recognition with RNN-transducer}, 
  year={2017},
  volume={},
  number={},
  pages={193-199},
  doi={10.1109/ASRU.2017.8268935}}

@InProceedings{pmlr-v48-amodei16,
  title = 	 {Deep Speech 2 : End-to-End Speech Recognition in English and Mandarin},
  author = 	 {Amodei, Dario and Ananthanarayanan, Sundaram and Anubhai, Rishita and Bai, Jingliang and Battenberg, Eric and Case, Carl and Casper, Jared and Catanzaro, Bryan and Cheng, Qiang and Chen, Guoliang and Chen, Jie and Chen, Jingdong and Chen, Zhijie and Chrzanowski, Mike and Coates, Adam and Diamos, Greg and Ding, Ke and Du, Niandong and Elsen, Erich and Engel, Jesse and Fang, Weiwei and Fan, Linxi and Fougner, Christopher and Gao, Liang and Gong, Caixia and Hannun, Awni and Han, Tony and Johannes, Lappi and Jiang, Bing and Ju, Cai and Jun, Billy and LeGresley, Patrick and Lin, Libby and Liu, Junjie and Liu, Yang and Li, Weigao and Li, Xiangang and Ma, Dongpeng and Narang, Sharan and Ng, Andrew and Ozair, Sherjil and Peng, Yiping and Prenger, Ryan and Qian, Sheng and Quan, Zongfeng and Raiman, Jonathan and Rao, Vinay and Satheesh, Sanjeev and Seetapun, David and Sengupta, Shubho and Srinet, Kavya and Sriram, Anuroop and Tang, Haiyuan and Tang, Liliang and Wang, Chong and Wang, Jidong and Wang, Kaifu and Wang, Yi and Wang, Zhijian and Wang, Zhiqian and Wu, Shuang and Wei, Likai and Xiao, Bo and Xie, Wen and Xie, Yan and Yogatama, Dani and Yuan, Bin and Zhan, Jun and Zhu, Zhenyao},
  booktitle = 	 {Proceedings of The 33rd International Conference on Machine Learning},
  pages = 	 {173--182},
  year = 	 {2016},
  editor = 	 {Balcan, Maria Florina and Weinberger, Kilian Q.},
  volume = 	 {48},
  series = 	 {Proceedings of Machine Learning Research},
  address = 	 {New York, New York, USA},
  month = 	 {20--22 Jun},
  publisher =    {PMLR},
  url = 	 {https://proceedings.mlr.press/v48/amodei16.html}
}

@article{graves2005framewise,
  title={Framewise phoneme classification with bidirectional LSTM and other neural network architectures},
  author={Graves, Alex and Schmidhuber, J{\"u}rgen},
  journal={Neural networks},
  volume={18},
  number={5-6},
  pages={602--610},
  year={2005},
  publisher={Elsevier}
}

@INPROCEEDINGS{7472621,
  author={Chan, William and Jaitly, Navdeep and Le, Quoc and Vinyals, Oriol},
  booktitle={2016 IEEE International Conference on Acoustics, Speech and Signal Processing (ICASSP)}, 
  title={Listen, attend and spell: A neural network for large vocabulary conversational speech recognition}, 
  year={2016},
  volume={},
  number={},
  pages={4960-4964},
  doi={10.1109/ICASSP.2016.7472621}}

@inproceedings{NIPS2017_3f5ee243,
 author = {Vaswani, Ashish and Shazeer, Noam and Parmar, Niki and Uszkoreit, Jakob and Jones, Llion and Gomez, Aidan N and Kaiser, \L ukasz and Polosukhin, Illia},
 booktitle = {Advances in Neural Information Processing Systems},
 editor = {I. Guyon and U. Von Luxburg and S. Bengio and H. Wallach and R. Fergus and S. Vishwanathan and R. Garnett},
 pages = {},
 publisher = {Curran Associates, Inc.},
 title = {Attention is All you Need},
 url = {https://proceedings.neurips.cc/paper_files/paper/2017/file/3f5ee243547dee91fbd053c1c4a845aa-Paper.pdf},
 volume = {30},
 year = {2017}
}

@INPROCEEDINGS{8462506,
  author={Dong, Linhao and Xu, Shuang and Xu, Bo},
  booktitle={2018 IEEE International Conference on Acoustics, Speech and Signal Processing (ICASSP)}, 
  title={Speech-Transformer: A No-Recurrence Sequence-to-Sequence Model for Speech Recognition}, 
  year={2018},
  volume={},
  number={},
  pages={5884-5888},
  doi={10.1109/ICASSP.2018.8462506}}

@InProceedings{pmlr-v162-baevski22a,
  title = 	 {data2vec: A General Framework for Self-supervised Learning in Speech, Vision and Language},
  author =       {Baevski, Alexei and Hsu, Wei-Ning and Xu, Qiantong and Babu, Arun and Gu, Jiatao and Auli, Michael},
  booktitle = 	 {Proceedings of the 39th International Conference on Machine Learning},
  pages = 	 {1298--1312},
  year = 	 {2022},
  editor = 	 {Chaudhuri, Kamalika and Jegelka, Stefanie and Song, Le and Szepesvari, Csaba and Niu, Gang and Sabato, Sivan},
  volume = 	 {162},
  series = 	 {Proceedings of Machine Learning Research},
  month = 	 {17--23 Jul},
  publisher =    {PMLR},
  url = 	 {https://proceedings.mlr.press/v162/baevski22a.html}
}

@InProceedings{pmlr-v202-radford23a,
  title = 	 {Robust Speech Recognition via Large-Scale Weak Supervision},
  author =       {Radford, Alec and Kim, Jong Wook and Xu, Tao and Brockman, Greg and Mcleavey, Christine and Sutskever, Ilya},
  booktitle = 	 {Proceedings of the 40th International Conference on Machine Learning},
  pages = 	 {28492--28518},
  year = 	 {2023},
  editor = 	 {Krause, Andreas and Brunskill, Emma and Cho, Kyunghyun and Engelhardt, Barbara and Sabato, Sivan and Scarlett, Jonathan},
  volume = 	 {202},
  series = 	 {Proceedings of Machine Learning Research},
  month = 	 {23--29 Jul},
  publisher =    {PMLR},
  url = 	 {https://proceedings.mlr.press/v202/radford23a.html}
}

@misc{gulati2020conformerconvolutionaugmentedtransformerspeech,
      title={Conformer: Convolution-augmented Transformer for Speech Recognition}, 
      author={Anmol Gulati and James Qin and Chung-Cheng Chiu and Niki Parmar and Yu Zhang and Jiahui Yu and Wei Han and Shibo Wang and Zhengdong Zhang and Yonghui Wu and Ruoming Pang},
      year={2020},
      eprint={2005.08100},
      archivePrefix={arXiv},
      primaryClass={eess.AS},
      url={https://arxiv.org/abs/2005.08100}, 
}

@online{Andrew_2025,
  author = {Andrew Seagraves},
  title = {Benchmarking Top Open Source Speech Recognition Models: Whisper, Facebook wav2vec2, and Kaldi},
  year = 2025,
  url = {https://deepgram.com/learn/benchmarking-top-open-source-speech-models},
  urldate = {2026-03-17}
}

@misc{yu2021fastemitlowlatencystreamingasr,
      title={FastEmit: Low-latency Streaming ASR with Sequence-level Emission Regularization}, 
      author={Jiahui Yu and Chung-Cheng Chiu and Bo Li and Shuo-yiin Chang and Tara N. Sainath and Yanzhang He and Arun Narayanan and Wei Han and Anmol Gulati and Yonghui Wu and Ruoming Pang},
      year={2021},
      eprint={2010.11148},
      archivePrefix={arXiv},
      primaryClass={eess.AS},
      url={https://arxiv.org/abs/2010.11148}, 
}

@misc{faster-whisper,
  author = {Klein, Guillaume and Ashraf, Mahmoud and others},
  title = {{Faster Whisper}: Transcription with {CTranslate2}},
  year = {2025},
  publisher = {GitHub},
  journal = {GitHub repository},
  howpublished = {\url{https://github.com/SYSTRAN/faster-whisper}},
  commit = {ed9a06cd89a93e47838f564998a6c09b655d7f43}
}

@INPROCEEDINGS{9291818,
  author={Leow, Chee Siang and Hayakawa, Tomoaki and Nishizaki, Hiromitsu and Kitaoka, Norihide},
  booktitle={2020 IEEE 9th Global Conference on Consumer Electronics (GCCE)}, 
  title={Development of a Low-Latency and Real-Time Automatic Speech Recognition System}, 
  year={2020},
  volume={},
  number={},
  pages={925-928},
  doi={10.1109/GCCE50665.2020.9291818}}

@misc{wong2024syllablebaseddnnhmmcantonese,
      title={Syllable based DNN-HMM Cantonese Speech to Text System}, 
      author={Timothy Wong and Claire Li and Sam Lam and Billy Chiu and Qin Lu and Minglei Li and Dan Xiong and Roy Shing Yu and Vincent T. Y. Ng},
      year={2024},
      eprint={2402.08788},
      archivePrefix={arXiv},
      primaryClass={cs.CL},
      url={https://arxiv.org/abs/2402.08788}, 
}

@misc{zhang2024conformer1robustasrlargescale,
      title={Conformer-1: Robust ASR via Large-Scale Semisupervised Bootstrapping}, 
      author={Kevin Zhang and Luka Chkhetiani and Francis McCann Ramirez and Yash Khare and Andrea Vanzo and Michael Liang and Sergio Ramirez Martin and Gabriel Oexle and Ruben Bousbib and Taufiquzzaman Peyash and Michael Nguyen and Dillon Pulliam and Domenic Donato},
      year={2024},
      eprint={2404.07341},
      archivePrefix={arXiv},
      primaryClass={eess.AS},
      url={https://arxiv.org/abs/2404.07341}, 
}

@misc{simic2023selfsupervisedadaptiveavfusion,
      title={Self-Supervised Adaptive AV Fusion Module for Pre-Trained ASR Models}, 
      author={Christopher Simic and Tobias Bocklet},
      year={2023},
      eprint={2312.13873},
      archivePrefix={arXiv},
      primaryClass={cs.SD},
      url={https://arxiv.org/abs/2312.13873}, 
}

@misc{wang2022wav2vecswitchcontrastivelearningoriginalnoisy,
      title={Wav2vec-Switch: Contrastive Learning from Original-noisy Speech Pairs for Robust Speech Recognition}, 
      author={Yiming Wang and Jinyu Li and Heming Wang and Yao Qian and Chengyi Wang and Yu Wu},
      year={2022},
      eprint={2110.04934},
      archivePrefix={arXiv},
      primaryClass={cs.CL},
      url={https://arxiv.org/abs/2110.04934}, 
}

@INPROCEEDINGS{10974078,
  author={Ojha, Shuubham and Gervits, Felix and Espy-Wilson, Carol},
  booktitle={2025 20th ACM/IEEE International Conference on Human-Robot Interaction (HRI)}, 
  title={Speaking with Robots in Noisy Environments}, 
  year={2025},
  volume={},
  number={},
  pages={1057-1061},
  doi={10.1109/HRI61500.2025.10974078}}

@INPROCEEDINGS{9414243,
  author={Subramanian, Aswin Shanmugam and Weng, Chao and Watanabe, Shinji and Yu, Meng and Xu, Yong and Zhang, Shi-Xiong and Yu, Dong},
  booktitle={ICASSP 2021 - 2021 IEEE International Conference on Acoustics, Speech and Signal Processing (ICASSP)}, 
  title={Directional ASR: A New Paradigm for E2E Multi-Speaker Speech Recognition with Source Localization}, 
  year={2021},
  volume={},
  number={},
  pages={8433-8437},
  doi={10.1109/ICASSP39728.2021.9414243}}

@inproceedings{MSDET,
author = {Hartanto, Roland and Sakti, Sakriani and Shinoda, Koichi},
year = {2024},
month = {09},
pages = {2170-2174},
title = {MSDET: Multitask Speaker Separation and Direction-of-Arrival Estimation Training},
doi = {10.21437/Interspeech.2024-2537}
}

@INPROCEEDINGS{9414187,
  author={Aroudi, Ali and Braun, Sebastian},
  booktitle={ICASSP 2021 - 2021 IEEE International Conference on Acoustics, Speech and Signal Processing (ICASSP)}, 
  title={DBnet: Doa-Driven Beamforming Network for end-to-end Reverberant Sound Source Separation}, 
  year={2021},
  volume={},
  number={},
  pages={211-215},
  doi={10.1109/ICASSP39728.2021.9414187}}

@ARTICLE{11329501,
  author={Xiong, Wenmeng and Jia, Maoshen and Zhou, Jing and Zhang, Jing and Shen, Qing},
  journal={IEEE Transactions on Audio, Speech and Language Processing}, 
  title={JointNet: Joint Learning for Simultaneous DOA Estimation and Speech Enhancement in Noisy and Reverberant Environments}, 
  year={2026},
  volume={34},
  number={},
  pages={596-611},
  doi={10.1109/TASLPRO.2026.3651053}}

@INPROCEEDINGS{10447153,
  author={Togami, Masahito and Valin, Jean-Marc and Helwani, Karim and Giri, Ritwik and Isik, Umut and Goodwin, Michael M.},
  booktitle={ICASSP 2024 - 2024 IEEE International Conference on Acoustics, Speech and Signal Processing (ICASSP)}, 
  title={Real-Time Stereo Speech Enhancement with Spatial-Cue Preservation Based on Dual-Path Structure}, 
  year={2024},
  volume={},
  number={},
  pages={71-75},
  doi={10.1109/ICASSP48485.2024.10447153}}

@Article{su132212392,
AUTHOR = {Gondi, Santosh and Pratap, Vineel},
TITLE = {Performance and Efficiency Evaluation of ASR Inference on the Edge},
JOURNAL = {Sustainability},
VOLUME = {13},
YEAR = {2021},
NUMBER = {22},
ARTICLE-NUMBER = {12392},
URL = {https://www.mdpi.com/2071-1050/13/22/12392},
ISSN = {2071-1050},
DOI = {10.3390/su132212392}
}

@article{Cherry1953Cocktail,
  author  = {Cherry, E. Colin},
  title   = {Some Experiments on the Recognition of Speech, with One and with Two Ears},
  journal = {The Journal of the Acoustical Society of America},
  volume  = {25},
  number  = {5},
  pages   = {975--979},
  year    = {1953},
  doi     = {10.1121/1.1907229}
}

@book{Bregman1990ASA,
    author = {Bregman, Albert S.},
    title = {Auditory Scene Analysis: The Perceptual Organization of Sound},
    publisher = {The MIT Press},
    year = {1990},
    month = {05},
    isbn = {9780262269209},
    doi = {10.7551/mitpress/1486.001.0001},
    url = {https://doi.org/10.7551/mitpress/1486.001.0001},
}

@article{Bronkhorst2000Cocktail,
  title={The cocktail party phenomenon: A review of research on speech intelligibility in multiple-talker conditions},
  author={Bronkhorst, Adelbert W},
  journal={ACUSTICA united with acta acustica},
  volume={86},
  number={1},
  pages={117--128},
  year={2000},
  publisher={S. Hirzel Verlag}
}

@article{HaebUmbach2021FarFieldASR,
  author  = {Haeb-Umbach, Reinhold and Heymann, Jahn and Drude, Lukas and Watanabe, Shinji and Delcroix, Marc and Nakatani, Tomohiro},
  title   = {Far-Field Automatic Speech Recognition},
  journal = {Proceedings of the IEEE},
  volume  = {109},
  number  = {2},
  pages   = {124--148},
  year    = {2021},
  doi     = {10.1109/JPROC.2020.3018668}
}

@misc{Barker2018CHiME5,
      title={The fifth 'CHiME' Speech Separation and Recognition Challenge: Dataset, task and baselines}, 
      author={Jon Barker and Shinji Watanabe and Emmanuel Vincent and Jan Trmal},
      year={2018},
      eprint={1803.10609},
      archivePrefix={arXiv},
      primaryClass={cs.SD},
      url={https://arxiv.org/abs/1803.10609}, 
}

@misc{Watanabe2020CHiME6,
      title={CHiME-6 Challenge:Tackling Multispeaker Speech Recognition for Unsegmented Recordings}, 
      author={Shinji Watanabe and Michael Mandel and Jon Barker and Emmanuel Vincent and Ashish Arora and Xuankai Chang and Sanjeev Khudanpur and Vimal Manohar and Daniel Povey and Desh Raj and David Snyder and Aswin Shanmugam Subramanian and Jan Trmal and Bar Ben Yair and Christoph Boeddeker and Zhaoheng Ni and Yusuke Fujita and Shota Horiguchi and Naoyuki Kanda and Takuya Yoshioka and Neville Ryant},
      year={2020},
      eprint={2004.09249},
      archivePrefix={arXiv},
      primaryClass={cs.SD},
      url={https://arxiv.org/abs/2004.09249}, 
}

@incollection{Doclo2010HearingAidBeamforming,
  title={Acoustic Beamforming for Hearing Aid Applications},
  author={Doclo, Simon and Gannot, Sharon and Moonen, Marc and Spriet, Ann},
  booktitle={Handbook on Array Processing and Sensor Networks},
  editor={Haykin, Simon and Liu, K. J. Ray},
  pages={269--302},
  publisher={Wiley},
  year={2010}
}

@article{Ahlawat2025ASRSurvey,
title = {Automatic Speech Recognition: A survey of deep learning techniques and approaches},
journal = {International Journal of Cognitive Computing in Engineering},
volume = {6},
pages = {201-237},
year = {2025},
issn = {2666-3074},
doi = {https://doi.org/10.1016/j.ijcce.2024.12.007},
url = {https://www.sciencedirect.com/science/article/pii/S2666307424000573},
author = {Harsh Ahlawat and Naveen Aggarwal and Deepti Gupta}
}

@inproceedings{machacek-etal-2023-turning,
    title = "Turning Whisper into Real-Time Transcription System",
    author = "Mach{\'a}{\v{c}}ek, Dominik  and
      Dabre, Raj  and
      Bojar, Ond{\v{r}}ej",
    editor = "Saha, Sriparna  and
      Sujaini, Herry",
    booktitle = "Proceedings of the 13th International Joint Conference on Natural Language Processing and the 3rd Conference of the Asia-Pacific Chapter of the Association for Computational Linguistics: System Demonstrations",
    month = nov,
    year = "2023",
    address = "Bali, Indonesia",
    publisher = "Association for Computational Linguistics",
    url = "https://aclanthology.org/2023.ijcnlp-demo.3",
    pages = "17--24",
}

@inproceedings{machacek-polak-2025-simultaneous,
    title = "Simultaneous Translation with Offline Speech and {LLM} Models in {CUNI} Submission to {IWSLT} 2025",
    author = "Mach{\'a}{\v{c}}ek, Dominik  and
      Pol{\'a}k, Peter",
    editor = "Salesky, Elizabeth  and
      Federico, Marcello  and
      Anastasopoulos, Antonis",
    booktitle = "Proceedings of the 22nd International Conference on Spoken Language Translation (IWSLT 2025)",
    month = jul,
    year = "2025",
    address = "Vienna, Austria (in-person and online)",
    publisher = "Association for Computational Linguistics",
    url = "https://aclanthology.org/2025.iwslt-1.41/",
    doi = "10.18653/v1/2025.iwslt-1.41",
    pages = "389--398",
    ISBN = "979-8-89176-272-5"
}

@article{van_den_bogaert2009speech,
    author = {Van den Bogaert, Tim and Doclo, Simon and Wouters, Jan and Moonen, Marc},
    title = {Speech enhancement with multichannel Wiener filter techniques in multimicrophone binaural hearing aids},
    journal = {The Journal of the Acoustical Society of America},
    volume = {125},
    number = {1},
    pages = {360-371},
    year = {2009},
    month = {01},
    issn = {0001-4966},
    doi = {10.1121/1.3023069},
    url = {https://doi.org/10.1121/1.3023069},
    eprint = {https://pubs.aip.org/asa/jasa/article-pdf/125/1/360/15289899/360_1_online.pdf},
}

@inproceedings{vandenbogaert2007binaural,
  title={Binaural cue preservation for hearing aids using an interaural transfer function multichannel Wiener filter},
  author={Van den Bogaert, Tim and Wouters, Jan and Doclo, Simon and Moonen, Marc},
  booktitle={2007 IEEE International Conference on Acoustics, Speech and Signal Processing-ICASSP'07},
  volume={4},
  pages={IV--565},
  year={2007},
  organization={IEEE}
}

@ARTICLE{marquardt2015interaural,
  author={Marquardt, Daniel and Hohmann, Volker and Doclo, Simon},
  journal={IEEE/ACM Transactions on Audio, Speech, and Language Processing}, 
  title={Interaural Coherence Preservation in Multi-Channel Wiener Filtering-Based Noise Reduction for Binaural Hearing Aids}, 
  year={2015},
  volume={23},
  number={12},
  pages={2162-2176},
  doi={10.1109/TASLP.2015.2471096}}

@INPROCEEDINGS{corey2019cooperative,
  author={Corey, Ryan M. and Skarha, Matthew D. and Singer, Andrew C.},
  booktitle={2019 IEEE 8th International Workshop on Computational Advances in Multi-Sensor Adaptive Processing (CAMSAP)}, 
  title={Cooperative Audio Source Separation and Enhancement Using Distributed Microphone Arrays and Wearable Devices}, 
  year={2019},
  volume={},
  number={},
  pages={296-300},
  doi={10.1109/CAMSAP45676.2019.9022475}}

@INPROCEEDINGS{corey2020binaural,
  author={Corey, Ryan M. and Singer, Andrew C.},
  booktitle={ICASSP 2020 - 2020 IEEE International Conference on Acoustics, Speech and Signal Processing (ICASSP)}, 
  title={Binaural Audio Source Remixing with Microphone Array Listening Devices}, 
  year={2020},
  volume={},
  number={},
  pages={561-565},
  doi={10.1109/ICASSP40776.2020.9053225}}

@article{argentieri2015survey,
title = {A survey on sound source localization in robotics: From binaural to array processing methods},
journal = {Computer Speech \& Language},
volume = {34},
number = {1},
pages = {87-112},
year = {2015},
issn = {0885-2308},
doi = {https://doi.org/10.1016/j.csl.2015.03.003},
url = {https://www.sciencedirect.com/science/article/pii/S0885230815000236},
author = {S. Argentieri and P. Danès and P. Souères}
}

@article{nakadai2010hark,
author = {Kazuhiro Nakadai and Toru Takahashi and Hiroshi G. Okuno and Hirofumi Nakajima and Yuji Hasegawa and Hiroshi Tsujino},
title = {Design and Implementation of Robot Audition System 'HARK' — Open Source Software for Listening to Three Simultaneous Speakers},
journal = {Advanced Robotics},
volume = {24},
number = {5-6},
pages = {739--761},
year = {2010},
publisher = {Taylor \& Francis},
doi = {10.1163/016918610X493561},


URL = { 
    
        https://doi.org/10.1163/016918610X493561
    
    

},
eprint = { 
    
        https://doi.org/10.1163/016918610X493561
    
    

}

}

@article{grondin2013manyears,
author = {Grondin, Fran\c{c}ois and L\'{e}tourneau, Dominic and Ferland, Fran\c{c}ois and Rousseau, Vincent and Michaud, Fran\c{c}ois},
title = {The ManyEars open framework},
year = {2013},
issue_date = {April 2013},
publisher = {Kluwer Academic Publishers},
address = {USA},
volume = {34},
number = {3},
issn = {0929-5593},
url = {https://doi.org/10.1007/s10514-012-9316-x},
doi = {10.1007/s10514-012-9316-x},
journal = {Auton. Robots},
month = apr,
pages = {217–232},
numpages = {16}
}

@Article{chhetri2018multichannel,
 author = {Amit S. Chhetri and Philip Hilmes and Trausti Kristjansson and Robert Ayrapetian and Wai Chu and Mohamed Mansour and Xiaoxue Li and Xianxian Zhang},
 title = {Multichannel audio front-end for far-field automatic speech recognition},
 year = {2018},
 url = {https://www.amazon.science/publications/multichannel-audio-front-end-for-far-field-automatic-speech-recognition},
}

@Article{iliev2023framework,
AUTHOR = {Iliev, Yuliy and Ilieva, Galina},
TITLE = {A Framework for Smart Home System with Voice Control Using NLP Methods},
JOURNAL = {Electronics},
VOLUME = {12},
YEAR = {2023},
NUMBER = {1},
ARTICLE-NUMBER = {116},
URL = {https://www.mdpi.com/2079-9292/12/1/116},
ISSN = {2079-9292},
DOI = {10.3390/electronics12010116}
}

@INPROCEEDINGS{foster2015chimehome,
  author={Foster, Peter and Sigtia, Siddharth and Krstulovic, Sacha and Barker, Jon and Plumbley, Mark D.},
  booktitle={2015 IEEE Workshop on Applications of Signal Processing to Audio and Acoustics (WASPAA)}, 
  title={Chime-home: A dataset for sound source recognition in a domestic environment}, 
  year={2015},
  volume={},
  number={},
  pages={1-5},
  doi={10.1109/WASPAA.2015.7336899}}

@inproceedings{carletta2005ami,
author = {Carletta, Jean and Ashby, Simone and Bourban, Sebastien and Flynn, Mike and Guillemot, Mael and Hain, Thomas and Kadlec, Jaroslav and Karaiskos, Vasilis and Kraaij, Wessel and Kronenthal, Melissa and Lathoud, Guillaume and Lincoln, Mike and Lisowska Masson, Agnes and Mccowan, Iain and Post, Wilfried and Reidsma, Dennis and Wellner, Pierre},
year = {2005},
month = {07},
pages = {},
title = {The AMI meeting corpus: A pre-announcement},
isbn = {978-3-540-32549-9},
journal = {Lecture Notes in Computer Science},
doi = {10.1007/11677482_3}
}

@misc{chen2020libricss,
      title={Continuous speech separation: dataset and analysis}, 
      author={Zhuo Chen and Takuya Yoshioka and Liang Lu and Tianyan Zhou and Zhong Meng and Yi Luo and Jian Wu and Xiong Xiao and Jinyu Li},
      year={2020},
      eprint={2001.11482},
      archivePrefix={arXiv},
      primaryClass={cs.SD},
      url={https://arxiv.org/abs/2001.11482}, 
}

@misc{heySiri2017,
  title={Hey Siri: An On-device DNN-powered Voice Trigger for Apple's Personal Assistant},
  author={{Apple Machine Learning Research}},
  year={2017},
  howpublished={Apple Machine Learning Journal},
  url={https://machinelearning.apple.com/research/hey-siri}
}

@article{ephrat2018looking,
author = {Ephrat, Ariel and Mosseri, Inbar and Lang, Oran and Dekel, Tali and Wilson, Kevin and Hassidim, Avinatan and Freeman, William T. and Rubinstein, Michael},
title = {Looking to listen at the cocktail party: a speaker-independent audio-visual model for speech separation},
year = {2018},
issue_date = {August 2018},
publisher = {Association for Computing Machinery},
address = {New York, NY, USA},
volume = {37},
number = {4},
issn = {0730-0301},
url = {https://doi.org/10.1145/3197517.3201357},
doi = {10.1145/3197517.3201357},
journal = {ACM Trans. Graph.},
month = jul,
articleno = {112},
numpages = {11}
}

@misc{turcotte2026lendearspeechenhancement,
      title={Lend me an Ear: Speech Enhancement Using a Robotic Arm with a Microphone Array}, 
      author={Zachary Turcotte and François Grondin},
      year={2026},
      eprint={2602.17818},
      archivePrefix={arXiv},
      primaryClass={cs.RO},
      url={https://arxiv.org/abs/2602.17818}, 
}

\vfill

\end{document}